\documentclass[reprint,nofootinbib,amsmath,amssymb,onecolumn,aps,11pt]{revtex4-2}
\usepackage{scrextend}
\usepackage{float}
\usepackage{appendix}
\usepackage{lipsum}
\usepackage[colorlinks, citecolor=blue,urlcolor=blue,linkcolor=blue,runcolor=filecolor]{hyperref}
\usepackage{url}
\usepackage{amsmath}
\usepackage{subfigure}
\allowdisplaybreaks[4]
\usepackage{graphicx}
\usepackage{dcolumn}
\usepackage{bm}
\begin{document}

\title{Investigating the environmental dependence of ultralight scalar dark matter with atom interferometers}
\author{Wei Zhao\textsuperscript{1,2}, Dongfeng Gao\textsuperscript{1,$^\ast$}, Jin Wang\textsuperscript{1,$\dagger$}, Mingsheng Zhan\textsuperscript{1,$\ddagger$}}
\address{\textsuperscript{1}State Key Laboratory of Magnetic Resonance and Atomic and Molecular Physics,
	\\Wuhan Institute of Physics and Mathematics, APM,\\
	Chinese Academy of Sciences, \\Wuhan 430071, China
	\\ \textsuperscript{2}School of Physical Sciences, \\University of Chinese Academy of Sciences, \\Beijing 100049, China
	\\$^\ast$dfgao@wipm.ac.cn\\$^\dagger$wangjin@wipm.ac.cn\\
	$^\ddagger$mszhan@wipm.ac.cn
}


\begin{abstract}
	We study the environmental dependence of ultralight scalar dark matter (DM) with linear interactions to the standard model particles. The solution to the DM field turns out to be a sum of the cosmic harmonic oscillation term and the local exponential fluctuation term. The amplitude of the first term depends on the local DM density and the mass of the DM field. The second term is induced by the local distribution of matter, such as the Earth. And it depends not only on the mass of the Earth, but also the density of the Earth. Then, we compute the phase shift induced by the DM field in atom interferometers (AIs), through solving the trajectories of atoms. Especially, the AI signal for the violation of weak equivalence principle (WEP) caused by the DM field is calculated. Depending on the values of the DM coupling parameters, contributions to the WEP violation from the first and second terms of the DM field can be either comparable or one larger than the other. Finally, we give some constraints to DM coupling parameters using results from the terrestrial atomic WEP tests.
\end{abstract}

\keywords{Atom interferometers, Ultralight dark matter, Weak equivalence principle}



\maketitle

\section{Introduction}\label{introduction}
A variety of astrophysical and cosmological observations indicate the existence of dark matter (DM) and dark energy, although we have not directly discovered them \cite{article,PhysRevLett.125.111101}. It is commonly believed that about 80\% of all the matter in the Universe is DM \cite{Abdullah2020}. So far the nature of DM is unknown except its gravitational effects at the galactic scale and larger \cite{doi:10.1146/annurev.astro.41.111302.102207,RN77,RN76}. There are considerable efforts to search for a kind of particle-like DM candidate---weakly interacting massive particle (WIMP). Unfortunately, no evidences of WIMP dark matter have been found \cite{PhysRevLett.118.021303, PhysRevLett.118.071301, PhysRevLett.122.071301}. In contrast, several experimental strategies are proposed recently to search for light, field-like DM using precision tools of atomic, molecular and optical physics, such as atomic clocks \cite{PhysRevD.91.015015,RN67}, atomic spectroscopy \cite{VanTilburg15PRL:AtomicSpectroscopyDM, RN37}, accelerometers \cite{RN45}, optical cavities \cite{Obata18,PhysRevLett.123.031304} and laser interferometers \cite{Stadnik15:LaserInterferometry,Stadnik16,Grote19PRR:GWDM}. 

In recent years, rapid technological progress in atom interferometry has been made. Atom interferometers (AIs) are realized by coherently manipulating atomic matter waves \cite{RevModPhys.81.1051}. The whole process mainly consists of preparing an atomic wave packet in the initial state, coherently splitting the wave packet into two by applying the laser pulse, flipping the atomic states of the two wave packets after some drift time $T$, recombining these wave packets after another drift time $T$, and finally measuring the phase shift of the detected fringes. AIs have already been used in various precision measurements. For example, the value of the fine structure constant was determined to be $\alpha^{-1}$ = 137.035999206(11) in the $^{87}$Rb-atom recoil experiment \cite{Nature588.61}, which is the most accurate measurement of $\alpha$ so far. AI has also been used to test weak equivalence principle (WEP) at quantum level. Recent results of quantum WEP test with AIs were reported by Zhou {\it et al.} \cite{zhou2019united} and Asenbaum {\it et al.} \cite{PhysRevLett.125.191101} with accuracies of $10^{-10}$-level and $10^{-12}$-level, respectively. 

Encouraged by the achievements AIs have made, people put forward several proposals of detecting ultralight DM with AIs \cite{RN45,PhysRevLett.117.261301,RN42}. The idea behind these proposals is the following. According to the popular scalar DM models \cite{RN18,Damour_2010}, the scalar DM may interact with standard-model matters and change the fundamental parameters, such as the mass of fermions, the electromagnetic fine structure constant and the QCD energy scale. This will lead to variations in atomic masses and atomic internal energy levels, and finally end up with a change in the mass of the Earth, resulting in a variation of the gravitational acceleration. All these effects can be searched by a net phase shift in AI experiments. But, in all these proposals, only the cosmic harmonic oscillation part of the DM field has been considered.

In this paper, we also work on the popular scalar DM models \cite{RN18, Damour_2010}. After a thorough computation, the solution to the DM field is obtained. The DM field is found to be a sum of the cosmic harmonic oscillation term and the local exponential fluctuation term. The second term comes from the local distribution of mass. We further calculate the signal for the WEP violation caused by the DM field in AIs. The calculation shows that contributions from the two terms of the DM field can be either comparable or one larger than the other, depending on the values of the DM coupling parameters.

The paper is organized as follows. In Sec. \ref{the scalar dark matter model}, the scalar DM model is briefly introduced. In Sec. \ref{the solution of the scalar field}, we discuss the environmental dependence of the scalar DM, and the solution to the DM field near a local distribution of matter (such as the Earth) is obtained. In Sec. \ref{shift phase}, we compute the phase shift in AI experiments under the influence of the scalar DM. In Sec. \ref{constraint the coupling parameters }, we discuss how to constrain the DM coupling parameters, using the newest atomic WEP tests. Finally, discussion and conclusion are made in Sec. \ref{conlusion and discussion}.

\section{The scalar dark matter model}\label{the scalar dark matter model}
In this section, we will briefly review the scalar DM model, introduced in Refs. \cite{RN18, Damour_2010}. The microscopic action of the model is the following,
\begin{align}
S &=\frac{1}{c}\int d^{4}x \dfrac{\sqrt{-g}}{2\kappa}\Bigg[R-2g^{\mu\nu}\partial_{\mu}\varphi \partial_{\nu}\varphi-V(\varphi)\Bigg]
\notag \\
&+\frac{1}{c}\int d^{4}x\sqrt{-g}\Bigg[\mathcal{L}_{SM}(g_{\mu\nu},\psi_{i})+\mathcal{L}_{int}(g_{\mu\nu},\varphi,\psi_{i})\Bigg]\, ,   \label{1}
\end{align}   
where $\kappa=\frac{8\pi G}{c^4}$. $R$ is the Ricci scalar of the spacetime metric $g_{\mu\nu}$, and $\varphi$ denotes the dimensionless scalar DM field. Note that a dimensionful scalar DM field $\Phi$ is also used in literature. The relation is $\sqrt{4\pi}\Phi=\varphi M_P$, where $M_P=1.2\times 10^{19}$ GeV is the Planck energy scale. The first line in Eq. (\ref{1}) describes the action for general relativity and the DM field, with $V(\varphi)$ being the potential term of $\varphi$. Here we only consider the quadratic mass term in the potential,
\begin{align}
V(\varphi)=2\,\frac{c^{2} m_{\varphi}^{2}}{\hslash^2}\varphi^2 \label{2}\, ,
\end{align}
where $m_{\varphi}$ is the mass of the DM field.
$\mathcal{L}_{SM} $ is the Lagrangian density of the standard-model fields $ \psi_{i}$ , and $\mathcal{L}_{int}$ is the interaction Lagrangian density between the DM field and standard-model fields. 

To be specific, we focus on the linear coupling model, 
\begin{align}
\mathcal{L}_{int}&=\varphi \Bigg[  \frac{d_{e}}{4 e^2}F_{\mu\nu}F^{\mu\nu}
-\frac{d_{g}\beta_{3}}{2g_{3}}F^A_{\mu\nu}F^{A\mu\nu}    - \sum_{i=e,u,d} (d_{m_{i}}+\gamma_{m_{i}}d_{g})m_{i}\bar{\psi}_{i}\psi_{i} \Bigg] \, , \label{3}
\end{align}
where $d_{e}$ and $d_{g} $ are the couplings to the $U(1)$ electromagnetic and $SU(3)$ gluonic field terms, respectively. $d_{m_{e}}$, $d_{m_{u}}$ and $d_{m_{d}}$ are the couplings to the masses of electron and quarks. $g_{3}$ is the QCD gauge coupling, and $\beta_{3}$ is the $\beta$-function for $g_{3}$. $m_{i}$ denotes the fermionic masses (electron and quarks), $\gamma_{m_{i}}$ is the anomalous dimension due to the renormalization-group running of the quark masses, and $\psi_{i}$ are the fermion spinors.

It is easy to find that, in the linear coupling model, the Lagrangian leads to the $\varphi$-dependence for the following five physical quantities,
\begin{align}
\alpha(\varphi)&=(1+d_{e}\varphi)\alpha 
\notag \\
\Lambda_{3}(\varphi)&=(1+d_{g}\varphi)\Lambda_{3} 
\notag \\
m_{i}(\varphi)&=(1+d_{m_{i}}\varphi)m_{i},\quad i=e,u,d  \label{4}
\end{align}
where $\alpha$ is the electromagnetic fine structure constant, and $\Lambda_3$ is the QCD energy scale. Then, the physical meaning of the five coupling parameters ($d_{e}$, $d_{g} $, $d_{m_{e}}$, $d_{m_{u}}$ and $d_{m_{d}}$) is very clear. They just introduce a linear $\varphi$-dependence to the corresponding physical quantities.

For later discussion, it is convenient to rewrite the masses of up and down quarks into the form of symmetric and antisymmetric combinations,
\begin{align}
\hat{m} =\frac{m_u+m_d}{2}, \,\,\,\, \ 
\delta m =m_d-m_u \, .
\end{align}
Their corresponding $\varphi$-dependence is 
\begin{align}
\hat{m}(\varphi) =(1+d_{\hat{m}}\varphi)\hat{m}, \,\,\,\, \ 
\delta m(\varphi) =(1+d_{\delta m}\varphi)\delta m \, ,
\end{align}
with
\begin{align}
d_{\hat{m}} =\frac{m_ud_{m_u}+m_dd_{m_d}}{m_u+m_d},  \ 
d_{\delta m} =\frac{m_dd_{m_d}-m_ud_{m_u}}{m_d-m_u} . \label{5}
\end{align}

From the action (\ref{1}), it is straight to derive the field equations for the spacetime metric $g_{\mu\nu}$ and $\varphi$, which are
\begin{equation}
{R}_{\mu\nu}=\kappa[T_{\mu\nu}-\frac{1}{2}g_{\mu\nu}T]+2\partial_{\mu}\varphi\partial_{\nu}\varphi+\frac{1}{2}g_{\mu\nu}V(\varphi) \label{6}     
\end{equation}
and
\begin{equation}
-\frac{1}{c^2}\ddot{\varphi}-\bigtriangleup\varphi =-\frac{\kappa}{2}\frac{\partial\mathcal{L}_{int}}{\partial\varphi}+\frac{V^{'}(\varphi)}{4}\, . \label{8}
\end{equation}
The stress-energy tensor $T_{\mu\nu}$ is defined by
\begin{equation}      
T_{\mu\nu}=-\frac{2}{\sqrt{-g}}\frac{\delta\sqrt{-g}\mathcal{L}_{mat}}{\delta g^{\mu\nu}}\, , \label{7}
\end{equation}
where $\mathcal{L}_{mat}$ denotes the Lagrangian for the matter source. 

To solve the above field equations, one needs to write down the phenomenological Lagrangian $\mathcal{L}_{mat}$ for matter, in the spirit of the microscopic action (\ref{1}). Since ordinary matter is made of atoms, which can be further decomposed into fundamental particles (photons, electrons, gluons and quarks), the problem is then reduced to write down a phenomenological Lagrangian for atoms. In Ref. \cite{Damour_1992}, such a phenomenological treatment of matter was developed, where the atom was modeled as a massive point particle. The phenomenological action for matter was written as
\begin{equation}
S_{mat}[g_{\mu\nu},\varphi]=-c^2\sum_{{\rm atom}} \int_{{\rm atom}} m_{A}(\varphi)\, d\tau \, , \label{9}
\end{equation}
where $\tau$ is the proper time along the atom's worldline, and $m_{A}$ is the atomic mass. Since each atom has its own decomposition, $ m_{A}(\varphi)$ has different dependence on $\varphi$.

In the paper \cite{RN18}, derived from the microscopic action (\ref{1}), a dimensionless phenomenological factor $\alpha_{A}$ is introduced to measure the coupling of DM field to the atom,
\begin{equation}
\alpha_{A}\equiv\dfrac{\partial \text{ln}m_{A}(\varphi)}{\partial \varphi} \, . \label{10}
\end{equation}
The expression for $\alpha_{A}$ has been derived,  
\begin{align}
\alpha_{A}=&d_g+[(d_{\hat{m}}-d_g)Q_{\hat{m}}+(d_{\delta m}-d_g)Q_{\delta m}
+(d_{m_e}-d_g)Q_{m_e}+d_eQ_e]        \, ,                            \label{11}
\end{align}
where the dilaton charges are given by
\begin{subequations}
	\begin{align}
	Q_{\hat m}&=F_A\Bigg[0.093-\frac{0.036}{A^{1/3}}-0.02\frac{(A-2Z)^2}{A^2} \label{q14a}
	 -1.4\times 10^{-4}\frac{Z(Z-1)}{A^{4/3}}\Bigg]\\ 
	Q_{\delta m}&=F_A\Bigg[0.0017\frac{A-2Z}{A}\Bigg] \\
	Q_{m_e}&=F_A\Bigg[5.5\times 10^{-4}\frac{Z}{A}\Bigg]\\
	Q_e&=F_A\Bigg[-1.4+8.2 \frac{Z}{A}+7.7 \frac{Z(Z-1)}{A^{4/3}}\Bigg]\times 10^{-4}
	\end{align}
and
\begin{equation}
F_A=A m_{amu}/m_A\, . \label{q14d}
\end{equation}
\end{subequations}
$Z$ is the atomic number, $A$ is the mass number of atoms, and $m_{amu}$ is the atomic mass unit. The factor $F_A=1+\mathcal{O}(10^{-4})$ can be replaced by one in the lowest approximation.

\section{The solution of the DM field near the Earth}\label{the solution of the scalar field}
To show the environmental dependence of scalar DM field, we need to solve the field equation for $\varphi$ near a distribution of ordinary matter, such as the Earth. For simplicity, we will regard the Earth as a spherically symmetric ball with radius $R_E$, density $\rho_E$, and mass $M_E=4\pi R_E^3 \rho_E/3$. According to Eq. (\ref{9}), the phenomenological action for the Earth is 
\begin{equation}
S_{E}= \frac{1}{c}\int  \mathcal{L}_{E}\, \sqrt{-g}\, d^4 x=-c \int \rho_{E}(\varphi)\, \sqrt{-g}\, d^4 x \, .
\end{equation}

Let us dwell on the $\varphi$-dependence of $M_E$, which comes from the $\varphi$-dependence of atoms. Since the five coupling parameters ($d_{e}$, $d_{g} $, $d_{m_{e}}$, $d_{\hat{m}}$ and $d_{\delta m}$) and $\varphi$ are assumed to be very small, we could do Taylor expansion in $\varphi$ for the mass $M_E$,
\begin{equation}
M_{E}(\varphi)=M_E \Bigg[1+\alpha_E\varphi+\tilde{\alpha_{E}}\varphi^2 + \mathcal{O}(\varphi^3)\Bigg]\,
\label{earthmass}
\end{equation} 
with 
\begin{align}
M_{E}\equiv M_{E}(\varphi)\Bigg\vert_{\varphi=0}, \,\,\,\, \alpha_{E}\equiv\dfrac{\partial \text{ln}M_{E}(\varphi)}{\partial \varphi}\Bigg\vert_{\varphi=0} \, .
\end{align}

Note that, unlike other papers (such as \cite{PhysRevD.98.064051}), we truncate the Taylor expansion of $M_E$ at the second order in $\varphi$. The physical meaning of the $\varphi^2$-term will be clear soon.

The calculation of $\alpha_{E}$ is as follows. The Earth is made of various elements, $49.83\%$ Oxygen, $15.19\%$ Iron, $15.14\%$ Magnesium, $14.23\%$ Silicon, $2.14\%$ Sulfur, $1.38\%$ Aluminum and $1\%$ Calcium \cite{RN100}. We first calculate the $\alpha_{A}$ for each element, using Eqs. (\ref{11}, \ref{q14a}-\ref{q14d}). Then, the $\alpha_{E}$ is given by taking the atomic average over the Earth's isotopic composition,
\begin{align}
\alpha_{E}=&d_g+[0.08(d_{\hat{m}}-d_g)+2.35\times 10^{-5}(d_{\delta m}-d_g)
\notag \\
&+2.71\times 10^{-4}(d_{m_e}-d_g)+1.71\times 10^{-3}d_e]    \notag \\   
=&0.92d_g+0.08d_{\hat{m}}+2.35\times 10^{-5}d_{\delta m}
\notag \\
&2.71\times 10^{-4}d_{m_e}+1.71\times 10^{-3}d_e        \, . \label{alphaE}
\end{align}
Note that, in Refs. \cite{RN18, Damour_2010, PhysRevD.98.064051}, $\alpha_{E}\backsimeq d_g^{*}=d_g+0.093(d_{\hat{m}}-d_g)+0.00027d_e+0.000275(d_{m_{e}}-d_{g})$ is used, which is the composition-independent part of the full $\alpha_{E}$. In this paper, we use the full $\alpha_{E}$ to study the effects from all the $d_i$'s.

According to the microscopic action (\ref{1}), there are only linear interactions between the DM field and the standard-model fields. Naively, one would expect that the Earth's mass only had linear-dependence on the scalar DM field. But, from the point view of effect field theory, once a model has linear couplings between scalar field and the Standard model fields, people will also have induced quadratic, cubic, quartic couplings, and so on. \footnote{To see this, it is easier to use a different DM coupling parameter notation $\{\Lambda_{\gamma},\Lambda_{g},\Lambda_{i}\}$. This notation is related to ours by $d_e=M_P/(4\pi \Lambda_{\gamma})$, $d_g=M_P/(4\pi \Lambda_{g})$, and $d_{m_i}=M_P/(4\pi \Lambda_{i})$. These $\Lambda_{\gamma, g,i}$ parameters have dimensions of energy, which can be regarded as UV cutoffs of some underlying theory. According to the spirit of effective field theory, terms, like $\sum_{n=2}^{\infty} C^F_{n} (\Phi/\Lambda_{\gamma})^n (F_{\mu\nu})^2$, $\sum_{n=2}^{\infty} C^g_{n} (\Phi/\Lambda_{g})^n (F_{\mu\nu}^{A})^{2}$, and $\sum_{n=2}^{\infty} C^i_{n} (\Phi/\Lambda_{i})^n m_i \bar{\psi}_i\psi_i$, will be generated. In the end, these terms produce the higher order $\varphi$-dependence in the Earth's mass (\ref{earthmass}). } The exact calculation is very tricky and lengthy, which is beyond the scope of this paper. The detailed form of $\tilde{\alpha_{E}}$ will be studied in our future work.

Based on the above discussion, a phenomenological action, describing $\varphi$ near the Earth, is given by 
\begin{equation}
S =\frac{1}{c}\int d^{4}x \dfrac{\sqrt{-g}}{2\kappa}\Bigg[R-2g^{\mu\nu}\partial_{\mu}\varphi \partial_{\nu}\varphi-V(\varphi)\Bigg]+S_{E}.  \label{earthaction}
\end{equation}
It is straightforward to write down the field equation for $\varphi$,
\begin{align}
-\frac{1}{c^2}\ddot{\varphi}+\triangledown^{2}\varphi =\frac{\kappa}{2}\rho_Ec^2\alpha_E+\dfrac{V'_{\text{eff}}}{4}\, ,  \label{eqphi1}
\end{align} 
with
\begin{align}
V_{\text{eff}}(\varphi)&=V(\varphi)+\kappa\rho_Ec^2 \tilde{\alpha_{E}}\varphi^2=2\frac{c^{2}}{\hslash^2}\Bigg(m_{\varphi}^{2}+\frac{1}{2}\kappa\rho_E \hslash^2 \tilde{\alpha_{E}}\Bigg)\varphi^2 \, . \label{effpotential}
\end{align}
Note that a Minkowski spacetime metric $g_{\mu\nu}={\rm diag}(-1,1,1,1)$ is assumed. It is clear that the potential for $\varphi$ is changed from $V(\varphi)$ to the effective potential $V_{\text{eff}}(\varphi)$, due to the appearance of the Earth.

To solve the field equation (\ref{eqphi1}), let us first consider the case without the Earth. Obviously, Eq. (\ref{eqphi1}) becomes
\begin{align}
-\frac{1}{c^2}\ddot{\varphi}+\triangledown^{2}\varphi =\dfrac{V'(\varphi)}{4}\, .  \label{eqphi2}
\end{align} 
It is easy to find out the solution,
\begin{equation}
\varphi_{bg}(t,{\textbf x})=\varphi_{0} \cos({\textbf k} \cdot {\textbf x}-\omega t+\delta)\, ,
\end{equation}  
where $\varphi_{0}$ is the amplitude, $\omega^2=\arrowvert{\textbf k}\arrowvert^2 c^2+m_{\varphi}^{2}c^4/\hslash^2 $, and $\delta$ is the initial phase. The solution is a plane wave, and we call it the background of $\varphi$. 

As in Ref. \cite{PhysRevD.91.015015}, the harmonic oscillation background $\varphi_{bg}$ will be identified as the DM. The wave vector is then given by ${\textbf k}=m_{\varphi} {\textbf v}_{{\rm vir}}/\hslash$, where $ {\textbf v}_{{\rm vir}}\backsimeq 10^{-3}c$ is the Earth's velocity with respect to the DM halo. Note that $\textbf v_{\rm vir}$ is seasonally modulated at a level of 10\% due to the Earth’s orbit around the Sun \cite{RN173}. For simplicity, we will ignore the modulation and take $\textbf k\cdotp \textbf x=kr$ in the following.

$\varphi_{bg}$ will contribute an energy density, $\frac{1}{8\pi}m_{\varphi}^{2}\varphi_{0}^2M_{Pl}^2$,  where $M_{Pl}=1.2\times 10^{19} {\rm GeV}$ is the Planck mass. Using the DM energy density $\rho_{DM} =0.4 {\rm GeV/cm^3}$ at the solar system, the amplitude is calculated to be
\begin{equation}
\varphi_{0} = \frac{7.2\times 10^{-31}{\rm eV}}{m_{\varphi}}\, .
\label{phi0}
\end{equation}  
It is easy to see that $\varphi_{0}\ll 1$ for the ultralight scalar DM ($10^{-22} {\rm eV}\lesssim m_{\varphi}\lesssim 1 {\rm eV}$). 

The appearance of the Earth will induce a perturbation to the background $\varphi_{bg}$. Let us decompose $\varphi$ into 
$$\varphi=\varphi_{bg}+\delta \varphi\, ,$$
where $\delta\varphi$ is the fluctuation around $\varphi_{bg}$. Insert it into Eq. (\ref{eqphi1}), we have  
\begin{align}
\triangledown^{2}\delta \varphi-\dfrac{c^2}{\hslash^2}m_{\text{eff}}\delta \varphi=\frac{\kappa}{2}\rho_Ec^2\alpha_E  \, ,  \label{19}
\end{align}
with
\begin{align}
m_{\text{eff}}^2=m^2_{\varphi}+\frac{4\pi G \hbar^2}{c^4}\rho_{E}\tilde{\alpha_{E}} \, . \label{effmass}
\end{align}
Now, it is clear that keeping the $\varphi^2$-term in $M_E$ results in a change in the mass of $\delta \varphi$ to the effective mass $m_{\text{eff}}$, which depends on the density $\rho_{E}$ and $\tilde{\alpha_{E}}$. \footnote{Note that the case here is different from the issue of naturalness. There, corrections to the mass of scalar field are generated by one-loop Feynman diagrams, which depend on some (arbitrary) UV cutoff scale. On the other hand, our correction in Eq. (\ref{effmass}) comes from the $\varphi^2$-dependence in the Earth's mass as discussed above. We will omit the issue of naturalness in this paper, and readers can refer to the papers \cite{PhysRevD.91.015015,Grote19PRR:GWDM} for further discussion.} Inserting the number $\rho_{E}=5.5 \times 10^3\, {\rm kg/m^3}$, one gets
\begin{align}
m_{\text{eff}}^2=m^2_{\varphi}+ (1.4\times 10^{-18} {\rm eV})^2\tilde{\alpha_{E}} \, . 
\label{effmass1}
\end{align}
Obviously, for the Earth, the difference between $m_{\text{eff}}$ and $m_{\varphi}$ can be neglected since $\tilde{\alpha_{E}}$ is very small.

Inside the Earth ($r \leqslant R_{E}$), the solution of Eq. (\ref{19}) is
\begin{align}
\delta\varphi&=-\alpha_{E}I(\frac{r}{\lambda_{\text{eff}}})\frac{GM_{E}}{c^2}\frac{ e^{-\frac{r}{\lambda_{\text{eff}}}}}{r}
-\frac{3\alpha_{E}GM_{E}\lambda_{\text{eff}}^2}{c^2R_{E}^3r}\Bigg[(r+\lambda_{\text{eff}})e^{-\frac{r}{\lambda_{\text{eff}}}}
\notag \\
&-(R_{E}+\lambda_{\text{eff}})e^{-\frac{R_E}{\lambda_{\text{eff}}}}\Bigg]\sinh(\frac{r}{\lambda_{\text{eff}}})\, , \label{deltaphi1}
\end{align}
where $ I(x)=3 \,\frac{x{\rm cosh}(x)-{\rm sinh}(x)}{x^3} $, ${\rm sinh}(x)=\frac{e^x-e^{-x}}{2}$ and ${\rm cosh}(x)=\frac{e^x+e^{-x}}{2}$. The effective wavelength of the DM field is defined to be $ \lambda_{\text{eff}}=\dfrac{\hbar}{m_{\text{eff}}c} $.

Outside the Earth ($r \geqslant R_{E}$), the solution of Eq. (\ref{19}) is
\begin{equation}
\delta\varphi=-\alpha_{E}I(\frac{R_E}{\lambda_{\text{eff}}})\frac{GM_{E}}{c^2}\frac{ e^{-\frac{r}{\lambda_{\text{eff}}}}}{r} \, . \label{deltaphi2}
\end{equation} 

So, in the neighborhood of the Earth where the terrestrial AI experiment is performed, the full solution to $\varphi$ is
\begin{equation}
\varphi=\varphi_{0} \cos(k\, r-\omega t+\delta)-\alpha_{E}I(\frac{R_{E}}{\lambda_{\text{eff}}})\frac{GM_{E}}{c^2}\frac{ e^{-\frac{r}{\lambda_{\text{eff}}}}}{r}\, . \label{24} 
\end{equation}  
Similar result was also obtained in the paper \cite{PhysRevD.98.064051}. The difference is that we have $\lambda_{\text{eff}}$ in the exponential term, instead of $\lambda_\varphi=\dfrac{\hbar}{m_{\varphi}c}$.

\section{The DM signal in atom interferometers}\label{shift phase} 
\begin{figure}[htbp]
	\centering
	\includegraphics[width=9cm,height=5cm]{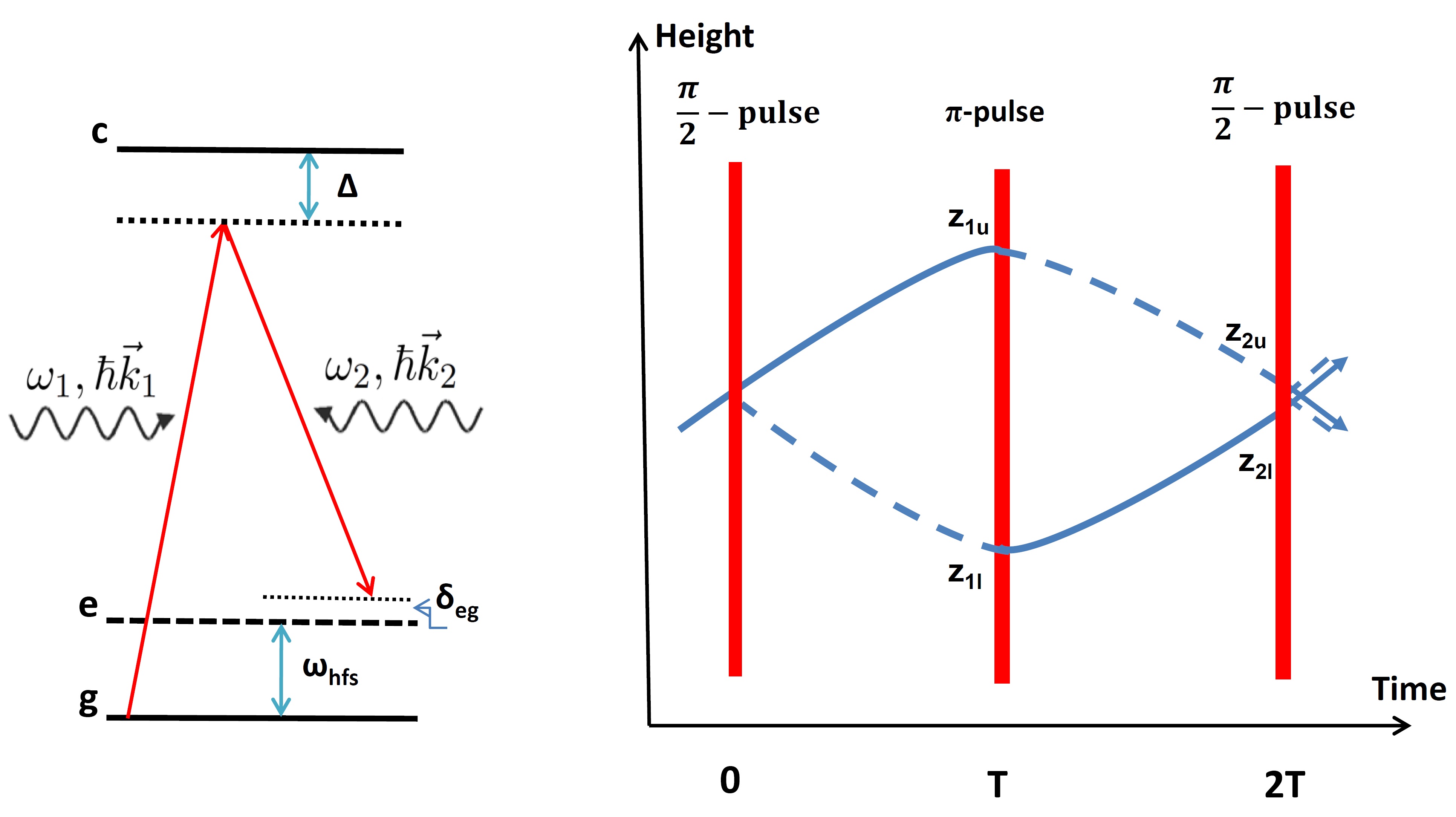}
	\caption{Schematic diagram for a typical $\frac{\pi}{2}$-$\pi$-$\frac{\pi}{2}$ Raman atom interferometer. The left part shows the stimulated Raman transition between two atomic hyperfine ground states $\vert g\rangle$ and $\vert e\rangle$. The atomic population is resonantly transferred between $\vert g\rangle$ and $\vert e\rangle$ if the frequency difference $\omega_1$-$\omega_2$ is close to $\omega_{{\rm hfs}}$. The right part shows the sequence of laser pulses and the paths of atoms.}
	\label{figure1}
\end{figure}

The theory of AIs can be found in many papers, such as Ref. \cite{ApplPhysB.54.321, PhysRevD.78.042003}. A typical $\frac{\pi}{2}$-$\pi$-$\frac{\pi}{2}$ Raman atom interferometer is shown in Fig. \ref{figure1}. The stimulated Raman transitions are realized by two counter-propagating laser beams. One is called the control laser beam, with frequency $\omega_1$ and wave vector ${\textbf k}_1$. The other one is called the passive laser beam, with frequency $\omega_2$ and wave vector ${\textbf k}_2$. The cold atom beam, prepared in the $\vert g\rangle$ state, is loaded into the AI with the launch velocity $v_L$.  At time t=0, the first Raman $\pi$/2-pulse is applied, and coherently splits the atomic wave packet into a superposition of states $\vert g\rangle$ and $\vert e\rangle$, with a momentum difference of ${\textbf k}_{{\rm eff}}$=${\textbf k}_1$-${\textbf k}_2$. After a drift time T, Raman $\pi$ pulses are applied, which transit the state $\vert g\rangle$ to $\vert e\rangle$, and the state $\vert e\rangle$ to $\vert g\rangle$, respectively. After another drift time T, the two wave packets overlap, and the final Raman  $\pi$/2-pulses are applied to make the two wave packets interfere. Then, the phase shift can be measured by detecting the number of atoms in either $\vert g\rangle$ or $\vert e\rangle$ states.

According to the discussion in previous sections, because of its interaction with standard-model fields, the scalar DM will change the fundamental parameters, such as the mass of fermions, the electromagnetic fine structure constant and the QCD energy scale. Subsequently, this will induce variations in the atomic masses, atomic internal energy levels, and the mass of the Earth, resulting in a variation of the gravitational acceleration. All these effects will cause a net phase shift in AI experiments, which signals the existence of the scalar DM. But, in previous proposals \cite{PhysRevLett.117.261301,RN42}, only the harmonic oscillation term of the DM field (\ref{24}) has been considered. The paper \cite{PhysRevLett.117.261301} considered the DM effects on the atomic masses and the mass of the Earth, while the authors in Ref. \cite{RN42} focused on the change in the atomic masses and atomic internal energy levels. In the following, we will give a complete computation for the phase shift due to variations in atomic masses, atomic internal energy levels, and the mass of the Earth, based on the full DM solution of $\varphi$. 

To calculate the DM-induced phase shift, we need to determine the atom's trajectories of the upper and lower arms, compute the phases along each arm, and take the phase difference between the two arms. The trajectory is determined by the atomic equation of motion, which can be derived from the non-relativistic approximation of the Lagrangian (\ref{9}),
\begin{equation}
L=-m_A c^2 +\frac{1}{2}m_{A}\dot{z}^2-m_{A}gz\, , \label{GAI3}
\end{equation} 
where $z\equiv r$-$R_E$ and $g$ is the Earth's gravitational acceleration. The DM-dependence of the Lagrangian (\ref{GAI3}) is encoded in $m_A$ and $g$. To be explicit, let us write down $m_A$ and $g$,
\begin{align}
m_{A}(\varphi)&=m_0(1+\alpha_{A}\varphi)=m_0\Bigg[1+\alpha_{A} \varphi_{0} \cos(kr-\omega t+\delta )\notag \\
&-\alpha_{A}\alpha_{E} I(\frac{R_{E}}{\lambda_{\text{eff}}})\frac{GM_{E}}{c^2}\frac{e^{-\frac{r}{\lambda_{\text{eff}}}}}{r}\Bigg]
\end{align}  
and 
\begin{align}
g(\varphi)&= GM_E(\varphi)/R_E^2=g_0\Bigg[1+\alpha_E \varphi_{0} \cos(kr-\omega t+\delta  )
\notag \\
&-\alpha_{E}^2 I(\frac{R_{E}}{\lambda_{\text{eff}}})\frac{GM_{E}}{c^2}\frac{ e^{-\frac{r}{\lambda_{\text{eff}}}}}{r}+ \mathcal{O}(\varphi_0^2)+ \mathcal{O}(d_i^3)\Bigg],
\end{align}
where $m_0$ and $g_0$ denote the atomic mass and the gravitational acceleration in the absence of DM, respectively. Higher order terms in $\varphi_0$ and $d_i$ are neglected. 

As pointed out in Ref. \cite{RN42}, we also need to consider the DM effect on atomic internal energy levels (i.e. $\vert c\rangle$, $\vert g\rangle$ and $\vert e\rangle$ in Fig. \ref{figure1}), which comes from the change in the electronic mass and the electromagnetic fine structure constant (\ref{4}). The change in atomic internal energy levels accordingly affects the stimulated atomic Raman transitions. In the end, the effective photon momentum transfer $k_{{\rm eff}}$ in stimulated atomic Raman transitions is changed to 
\begin{align}
k_{{\rm eff}}(\varphi)&=k_{{\rm eff}}\Bigg[1+(d_{m_{e}}+\xi d_{e})\varphi\Bigg]=k_{{\rm eff}}\Bigg[1+(d_{m_{e}}+\xi d_{e})\varphi_{0}\cos(kr-\omega t +\delta)
\notag \\
&-(d_{m_{e}}+\xi d_{e})\alpha_{E} I(\frac{R_{E}}{\lambda_{\text{eff}}})\frac{GM_{E}}{c^2}\frac{ e^{-\frac{r}{\lambda_{\text{eff}}}}}{r}\Bigg] ,
\label{equkeff}
\end{align}
where $k_{{\rm eff}}$ denotes the unperturbed value, and $\xi$ (=2.34 for the Rb atom) is the relativistic correction factor $2+K_{{\rm rel}}$ given in Ref. \cite{PhysRevC.73.055501}. Then, through the laser pulse's interaction with atoms, this effect is finally transferred to the atomic recoil velocity 
\begin{align}
v_{R}(\varphi)&=\frac{\hbar k_{\text{eff}}(\varphi)}{m_{A}(\varphi)} \backsimeq v_{R}\Bigg[1-(\alpha_{A}-\tilde{d})\varphi_{0} \cos(kr-\omega t+\delta)
\notag \\
&+(\alpha_{A}-\tilde{d})\alpha_{E} I(\frac{R_{E}}{\lambda_{\text{eff}}})\frac{GM_{E}}{c^2}\frac{e^{-\frac{r}{\lambda_{\text{eff}}}}}{r}\Bigg] ,
\end{align}
where  $v_{R}=\hbar k_{\text{eff}}/m_{0}$ denotes the unperturbed value, and $\tilde{d} \equiv d_{m_{e}}+\xi d_{e}$.

Now, it is straightforward to write down the atomic equation of motion from the Lagrangian (\ref{GAI3})
\begin{align}
m_{A}\ddot{z}&=-\frac{\partial m_{A}}{\partial z}c^2+\frac{1}{2}\frac{\partial m_{A}}{\partial z}\dot{z}^2-\dot{m}_{A}\dot{z}-\frac{\partial m_{A}}{\partial z}gz	-m_{A}\frac{\partial g}{\partial z}z -m_{A}g\, . \label{eom}
\end{align}
Solving Eq. (\ref{eom}) is very lengthy, and the full result will be given in Eqs.(\ref{zresult1}, \ref{zresult2}) in Appendix \ref{AppendixA}.

The total phase shift can be written as a sum of three components \cite{PhysRevD.78.042003}, the propagation phase shift, the laser phase shift, and the separation phase shift, 
\begin{equation}
\Delta \phi=\Delta\phi_{prop}+\Delta\phi_{laser}+\Delta\phi_{sep} \, .
\end{equation}
The calculation of the total phase shift is quite long, and will be given in Eqs. (\ref{phaseresult1}-\ref{phaseresult3}) in Appendix \ref{AppendixB}. 

To show that our result is a complete result, let us discuss two cases. First, consider effects induced by the local exponential fluctuation term $\delta\varphi$, which turns out to be the $\Delta\phi_{\delta\varphi}$ term (\ref{phaseresult2}) of the total phase shift $\Delta \phi$. In the $\lambda_{\text {eff}} \to \infty$ limit, the $\Delta\phi_{\delta\varphi}$ is reduced to
\begin{align}
& \tilde{\Delta\phi_{\delta\varphi}} =-g_{0}T^2 k_{\text{eff}}\left[1+\left(1+\frac{v_{L}(v_{R}+v_{L})}{2c^2}\right)\alpha_A\alpha_{E}\right]
\notag \\
& -g_{0}T^2 k_{\text{eff}}\left[ -\frac{7\,g^2_{0}\,{T}^{2}}{6c^2}+\frac{g_{0}(2\,v_{L}+v_{R})T-g_{0}R_{E}}{c^2}\right]\alpha_{E}^2 \,.
\end{align}
Then, it is easy to find that the E\"{o}tv\"{o}s parameter could be written as
\begin{align}
\tilde{\eta}_{\delta\varphi}\simeq(\alpha_{a}-\alpha_{b})\alpha_{E}
-\frac{ v_{L}(v_{R}+v_{L})}{2c^2}(\alpha_{a}-\alpha_{b})\alpha_{E}\, , \label{eta1}
\end{align}
for atomic species $a$ and $b$.
The first term in Eq. (\ref{eta1}) exactly reproduces the formula used in Ref. \cite{PhysRevLett.120.141101}. The second term in Eq. (\ref{eta1}) gives small corrections. For cold atoms, the velocity $v_L$ is about several ${\rm m/s}$. Thus, the corrections are about $10^{-17}$ times smaller than the first line. For hot atoms, the corrections can be much bigger.

The second case is to focus on effects caused by the the cosmic harmonic oscillation term $\varphi_{bg}$, which finally contributes the $\Delta\phi_{bg}$ term (\ref{phaseresult3}) in the total phase shift. If we ignore terms originated from the $m c^2$ term in the Lagrangian (\ref{GAI3}) and omit terms involving $v_{L,R}\!/\!c$ or $v_{L,R}T\!/\!R_E$, the $\Delta\phi_{bg}$ is reduced to 
\begin{align}
\tilde{\Delta\phi_{bg}} &=-\alpha_{A}\frac{2g_{0}k_{\text{eff}}T}{\omega}\varphi_{0}(\sin\omega T-\sin2\omega T)
\notag \\
&+(\alpha_{E}+2\alpha_{A}) \frac{g_{0}k_{\text{eff}}}{\omega^{2}}\varphi_{0}(1-2\cos\omega T+\cos2\omega T)
\notag \\
&+\alpha_{A}(\frac{k_{\text{eff}}(v_{L}+v_{R}/2)}{\omega})\varphi_{0}(2\sin\omega T-\sin 2\omega T)
\end{align}
One can see that we reproduce the result used in Ref. \cite{PhysRevLett.117.261301}.

\section{Constraints on the coupling parameters }\label{constraint the coupling parameters }

In this section, we discuss how to constrain the five coupling parameters ($d_{e}$, $d_{g} $, $d_{m_{e}}$, $d_{\hat{m}}$ and $d_{\delta m}$) by recent results of quantum WEP test with $^{85}$Rb-$^{87}$Rb duel-species AIs. $\eta$=$(-0.6\pm 3.7)\times10^{-10}$ was reported in Ref. \cite{zhou2019united} and $\eta$=$(1.6\pm 3.8)\times10^{-12}$ was obtained in Ref. \cite{PhysRevLett.125.191101}. For comparison, $\eta$=$(-1\pm 12.7)\times10^{-15}$, measured by the MICROSCOPE space mission \cite{PhysRevLett.119.231101} using macroscopic Ti-Pt objects, is also discussed. 

Since $v_{L,R} \ll c$, we can omit all the terms involving $v_{L,R}/c$ in Eq. (\ref{phaseresult1}-\ref{phaseresult3}). Then, the full result for $\Delta\phi$ can be simplified into 
	\begin{align}
	\Delta\phi&\backsimeq -g_{0}T^2k_{\text{eff}}-k_{\text{eff}}\frac{c^2k\alpha_{A}\varphi_{0}}{\omega^2}\Bigg[\sin(kR_{E}-2\omega T+\delta)-2\sin(kR_{E}-\omega T+\delta)\notag \\
	&+\sin(kR_{E}+\delta)\Bigg]
	+\alpha_{A}\frac{2g_{0}k_{\text{eff}}T}{\omega}\varphi_{0}\Bigg[\sin(kR_{E}-\omega T+\delta)-\sin(kR_{E}-2\omega T\notag \\
	&+\delta)\Bigg]
	+\left( \alpha_{E}+2\alpha_{A}\right) \frac{g_{0}k_{\text{eff}}}{\omega^{2}}\varphi_{0}\Bigg[\cos(kR_{E}+\delta)-2\cos(kR_{E}-\omega T+\delta)\notag \\
	&+\cos(kR_{E}-2\omega T+\delta)\Bigg]
	\notag \\
	&-T^2k_{\text{eff}}\Bigg[\frac{\frac{7}{6}g_{0}T^2-(2v_{L}+v_{R})T}{\lambda_{\varphi}}+(1+\frac{R_{E}}{\lambda_{\varphi}})\Bigg]I(\frac{R_{E}}{\lambda_{\varphi}})\frac{GM_{E}}{R_{E}^2}\alpha_{A}\alpha_{E}e^{-\frac{R_{E}}{\lambda_{\varphi}}}\, ,
	\end{align}
where we have replaced $\lambda_{\text{eff}}$ by $\lambda_{\varphi}$ since the difference is very small for the Earth, according to Eq. (\ref{effmass1}).

For $^{85}$Rb and $^{87}$Rb atoms, the DM-induced acceleration of gravity is given by	$-\frac{\Delta\phi}{T^2k_{\text{eff}}}$. Accordingly, the E$\ddot{\text{o}}$tv$\ddot{\text{o}}$s parameter is defined to be
	\begin{equation}
	\eta \equiv 2\frac{g_{85}-g_{87}}{g_{85}+g_{87}} \, .
	\end{equation}	
We find that $\eta$ can be written as a sum of a static component $\eta_{\delta\varphi}$ and an oscillatory component $\eta_{bg}$.
	\begin{equation}
	\eta = \eta_{\delta\varphi}+	\eta_{bg} \, .
	\end{equation}	
The $\delta\varphi$-contribution to $\eta$ is given by
	\begin{align}
	\eta_{\delta\varphi}= &\Bigg[\frac{\frac{7}{6}g_{0}T^2-(2v_{L}+v_{R})T}{\lambda_{\varphi}}+(1+\frac{R_{E}}{\lambda_{\varphi}})\Bigg]I(\frac{R_{E}}{\lambda_{\varphi}})(\alpha_{85}-\alpha_{87})\alpha_{E}e^{-\frac{R_{E}}{\lambda_{\varphi}}}\, ,
	\label{etadelphi}
	\end{align}
where
	\begin{align}
	\alpha_{85}&=9.1556222\times 10^{-1} d_{g}+8.3978\times 10^{-2}d_{\hat{m}}
+2.20\times 10^{-4}d_{\delta m} \notag \\
&+2.394 \times 10^{-4}d_{m_{e}}
	+2.961\times 10^{-3} d_{e}\, . 
	\end{align}
	\begin{align}
	\alpha_{87}&=9.1556685\times 10^{-1} d_{g}+8.3945\times 10^{-2}d_{\hat{m}}
	+2.54\times 10^{-4}d_{\delta m} \notag \\
	&+2.339 \times 10^{-4}d_{m_{e}}
	+2.869\times 10^{-3} d_{e}\, . 
	\end{align}
The $\varphi_{bg}$-contribution to $\eta$ is given by
	\begin{align}
	\eta_{bg}&=\!\Bigg[\frac{c^2k}{g_{0}\omega^2T^2}\Bigg(\sin(kR_{E}-2\omega T+\delta)\!-\!2\sin(kR_{E}-\omega T+\delta)\!+\!\sin(kR_{E}+\delta)\Bigg)
	\notag \\
	&-\frac{2}{T\omega}\Bigg(\sin(kR_{E}-\omega T+\delta)-\sin(kR_{E}-2\omega T+\delta)\Bigg)
	\notag \\
	&-  \frac{2}{\omega^{2}T^2}\Bigg(\cos(kR_{E}+\delta)-2\cos(kR_{E}-\omega T+\delta)+\cos(kR_{E}-2\omega T+\delta)\Bigg)\Bigg]
	\notag \\
	&\cdot(\alpha_{85}-\alpha_{87})\varphi_{0}
	\notag \\
	&=F(\delta,m_{\varphi},\varphi_{0})(\alpha_{85}-\alpha_{87})
	\, .
	\label{etabg}
	\end{align}
		\begin{figure}[H]
			\centering
			\includegraphics[width=9cm,height=6.75cm]{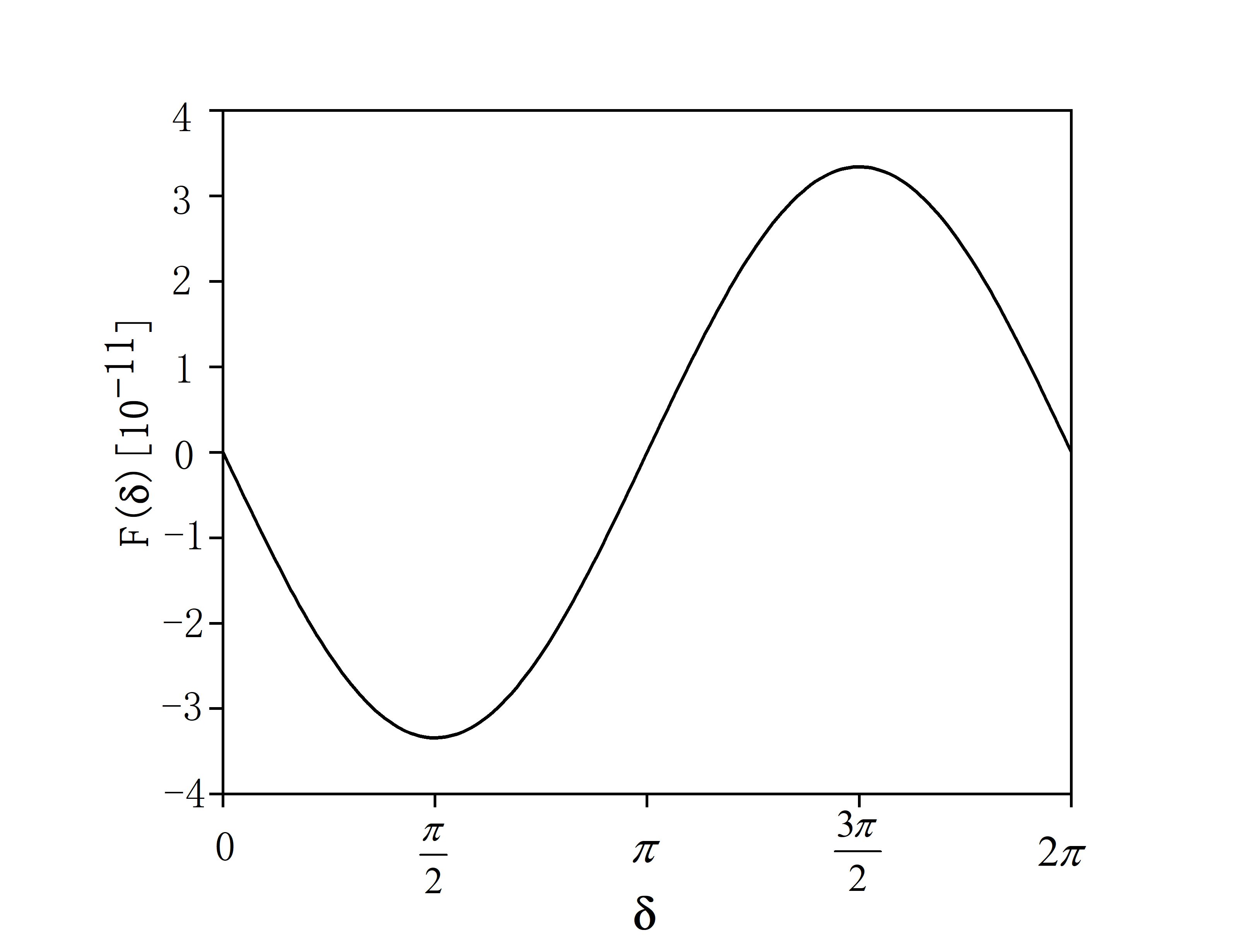}
			\caption{$\eta_{bg}$ is oscillatory in $\delta$, where $m_{\varphi}$ is taken to be $10^{-18}\rm {eV}$ for example.} \label{oscillation}
		\end{figure}
Note that $\eta_{bg}$ is oscillatory because $\delta$ (the initial phase of the DM field $\varphi_{bg}$) is different for each run of detection, as depicted in the Fig. \ref{oscillation}. To marginalize
the unknown initial phase $\delta$, we treat it as a random variable with a probability density function  
	\begin{align}
f(\delta)=\begin{cases}
	\frac{1}{2\pi} \quad 0<\delta<2\pi,\\
	\:0  \quad  \,\, {\rm otherwise} \, .
\end{cases}	
	\end{align}
Then, $\eta_{bg}$ becomes a function of the random variable $\delta$. According to the theory of probability and statistics (e.g. Ref. \cite{1997Schaum}), the mean (or expectation) of $\eta_{bg}$ is defined to be
\begin{align}
E[\eta_{bg}]=\int_{-\infty}^{+\infty}\eta_{bg}f(\delta)d\delta\,.
\end{align} 
It is easy to see that $E[\eta_{bg}]$=0.  So, for marginalizing $\delta$, we use the square root of $E[\eta_{bg}^2]$, which is 
	\begin{align}
	\bar{\eta}_{bg}
	&=\sqrt{\int_{-\infty}^{+\infty}\eta_{bg}^2 f(\delta)d\delta}
	\notag \\
	&=\frac{\sqrt{2}}{{g_{0}}{\omega}^{2}{T}^{2}}\, \bigg[\bigg( \cos  \left(\omega T\right)  -1 \bigg)  \bigg(\Big( 2\,g_{0}\,
	\omega\,T{c}^{2}k+{c}^{4}{k}^{2}+4\,{g_{0}}^{2}\Big)\Big(\cos \left(\omega T\right) -1\Big)
	\notag \\
	&+	2\,{g_{0}}^{2}\omega\,T\Big(2\sin \left(
	\omega T\right)-\omega\,T\Big)
	  \bigg) \bigg]^{1/2}\, \big(\alpha_{85}-\alpha_{87}\big) \, \varphi_{0}\,.
	  \label{signal amplitude}
	\end{align}
	The other notable thing is that $\eta_{bg}$ is linearly proportional to the $d_i$ parameters, while $\eta_{\delta\varphi}$ is quadratic in them. Then, $\eta_{\delta\varphi}$ and $\bar{\eta}_{bg}$ can be either comparable or one larger than the other depending on the values of $d_i$'s.

\begin{figure}[H]
	\centering
	\includegraphics[width=0.49\textwidth]{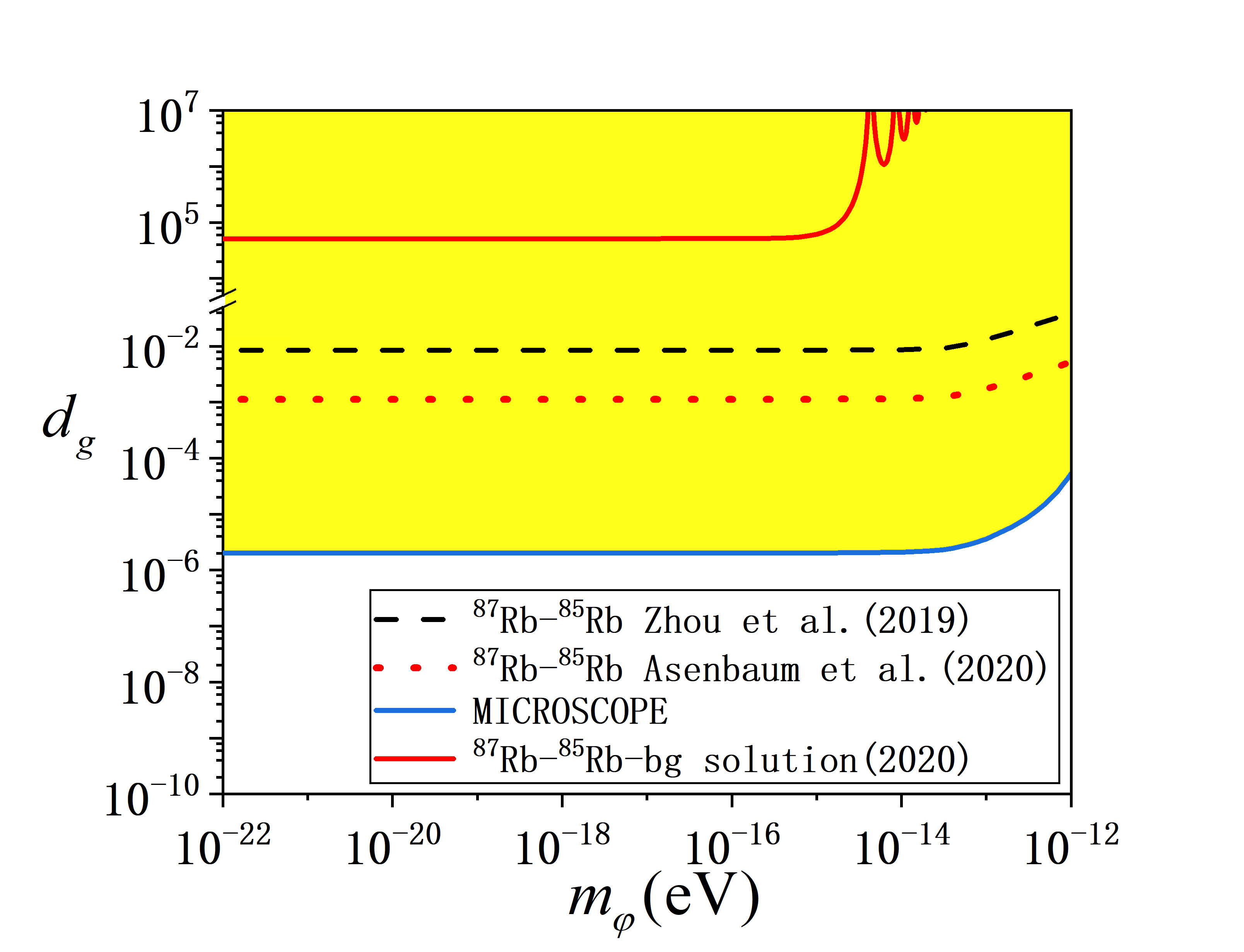}
	\vspace{0.5cm}
	\includegraphics[width=0.49\textwidth]{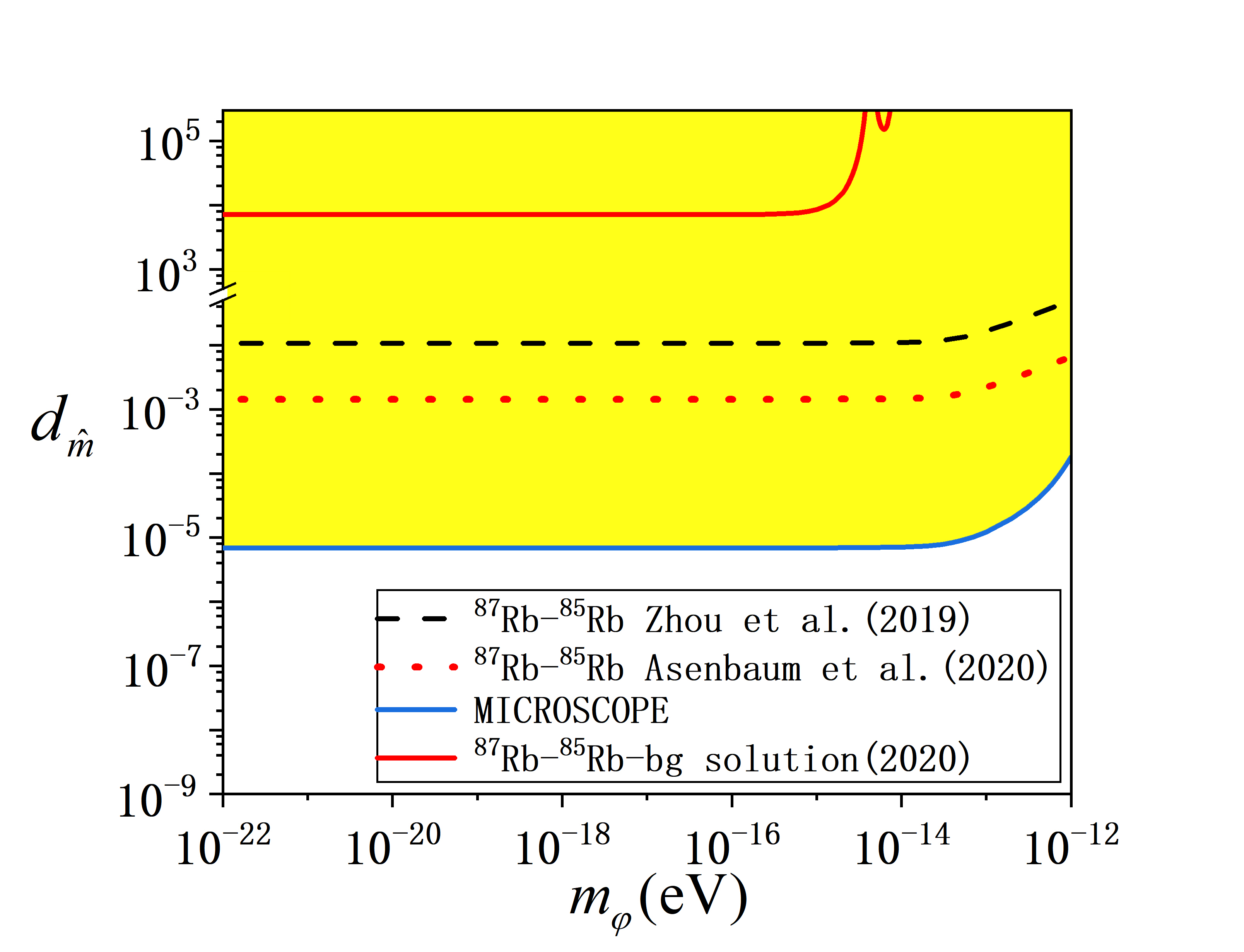}
	\includegraphics[width=0.49\textwidth]{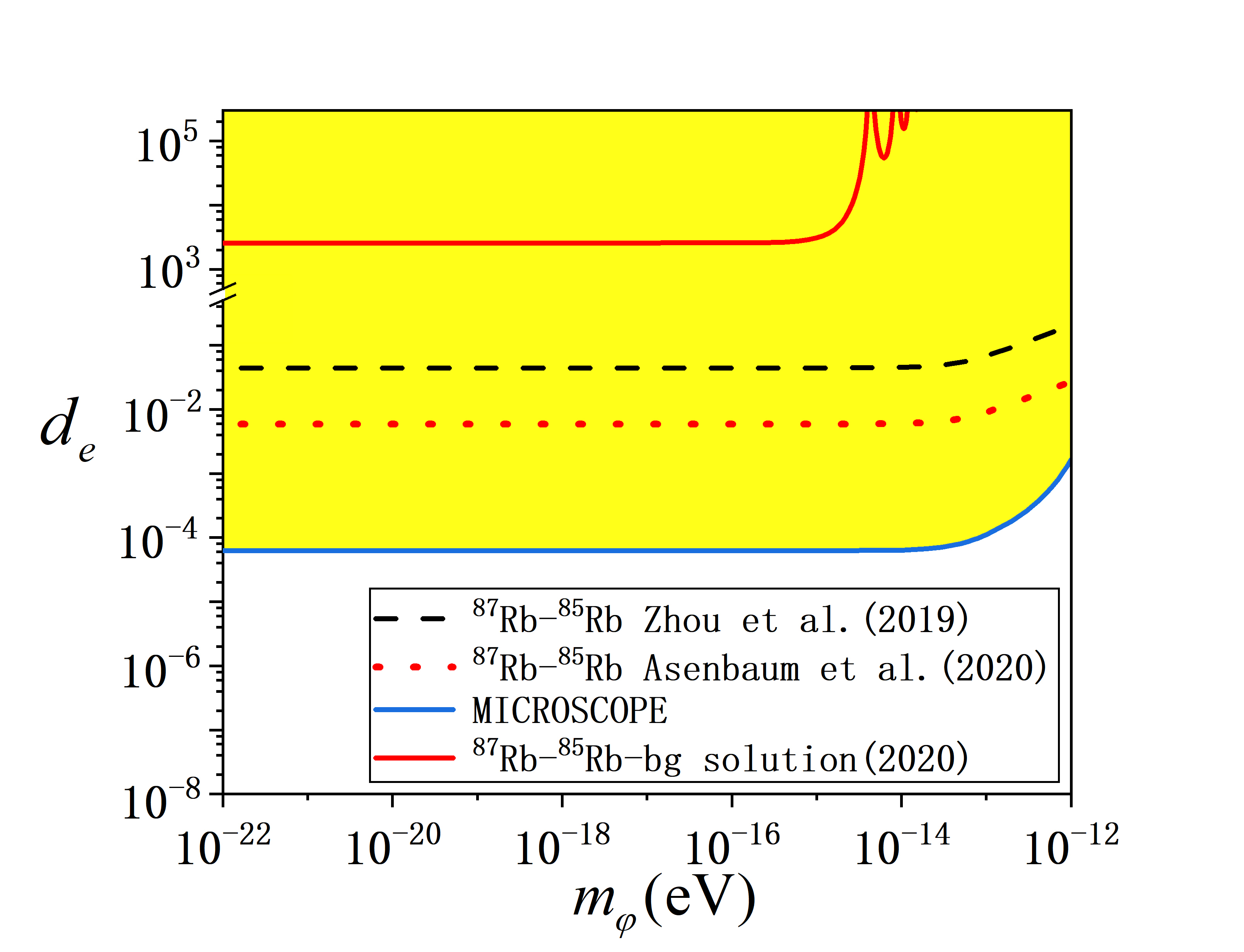}
	\vspace{0.5cm}
	\includegraphics[width=0.49\textwidth]{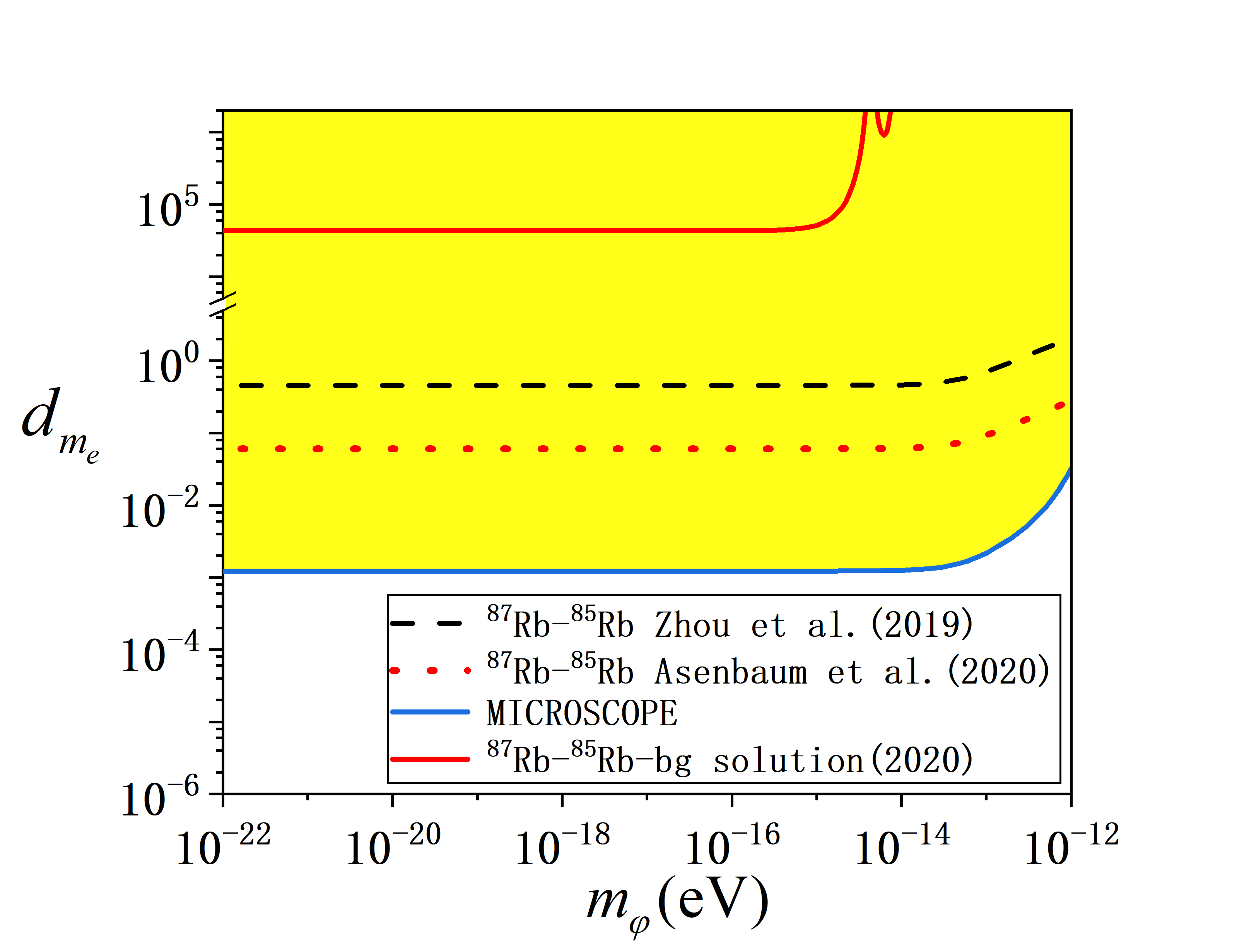} 
	\includegraphics[width=0.49\textwidth]{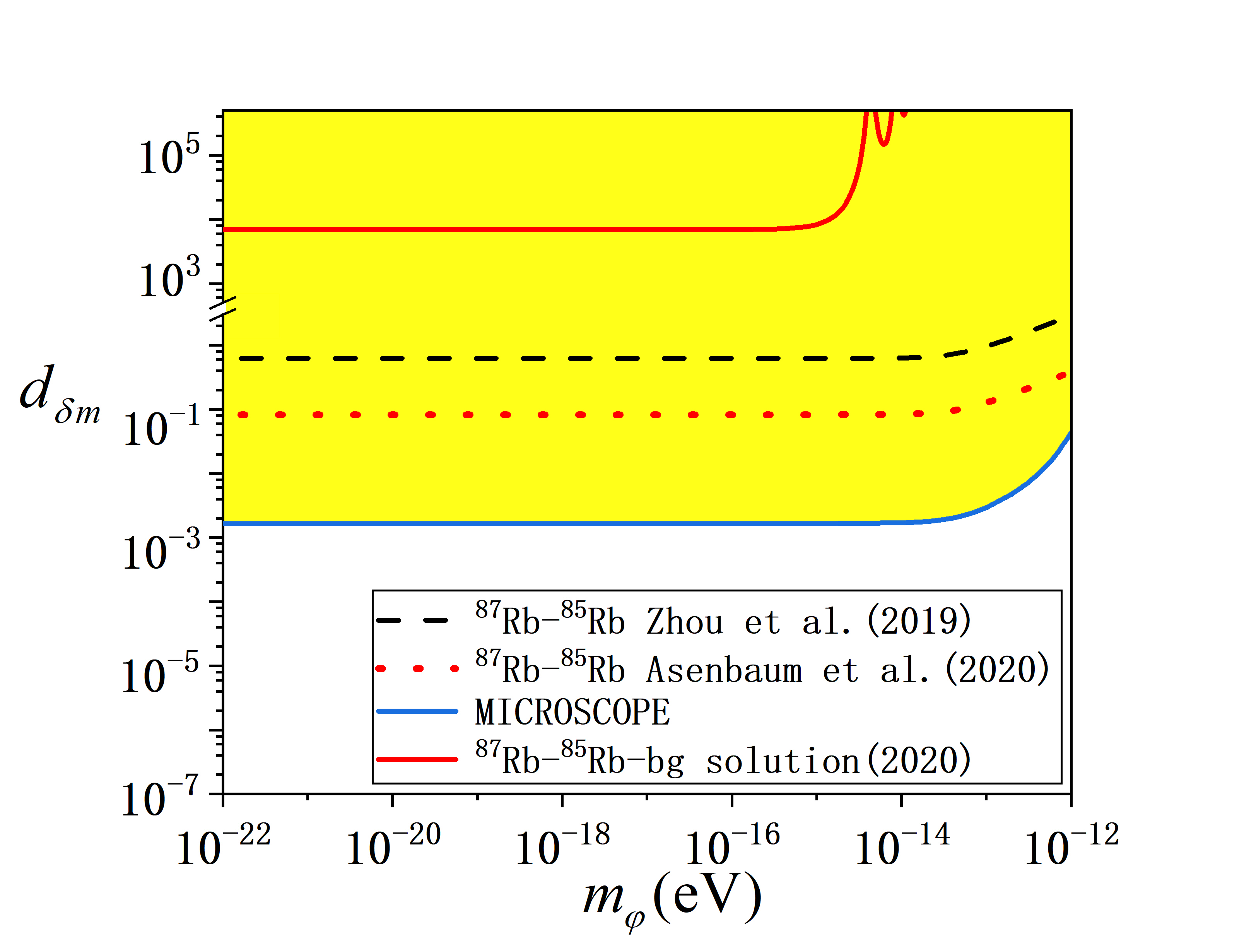}
	\caption{Constraints on the five DM parameters, $d_{e}$, $d_{g} $, $d_{m_{e}}$, $d_{\hat{m}}$ and $d_{\delta m}$. The blue solid line (and the corresponding yellow shaded area) is the constraint set by the MICROSCOPE's result \cite{PhysRevLett.119.231101}. The red dot-dashed line is the constraint set by Asenbaum's result \cite{PhysRevLett.125.191101} and the black dotted line is the constraint set by Zhou's result \cite{zhou2019united}. In addition, the red solid line is the constraint set by Asenbaum's result, considering only the component $\bar{\eta}_{bg}$.}
	\label{fig2}	
\end{figure}

	We first constrain only one parameter each time with the other four parameters set to zero. This method is widely used in many papers \cite{RN45,PhysRevD.98.064051,PhysRevLett.120.141101}. The result is shown in Fig. \ref{fig2}. It is clear that the MICROSCOPE's result gives better constraints on all five parameters than AI experiments since it keeps the best precision on WEP test. In Fig. \ref{fig2}, we can see that constraints on $d_{g} $ and $d_{\hat{m}}$ are better than constraints on the other three parameters. As explained in Refs. \cite{RN18, Damour_2010}, the reason is because the gluon interaction (i.e. the strong interaction) and quark masses make the most contribution to the atomic masses, which can be seen from the coefficients of $d_i$'s in Eqs. (\ref{11}) and (\ref{alphaE}). At current precision level on WEP test, the oscillatory component $\eta_{bg}$ makes neglectable contributions to constraints on the five DM parameters. 
	
\begin{figure}[H]	
	\centering
	\includegraphics[width=0.49\textwidth]{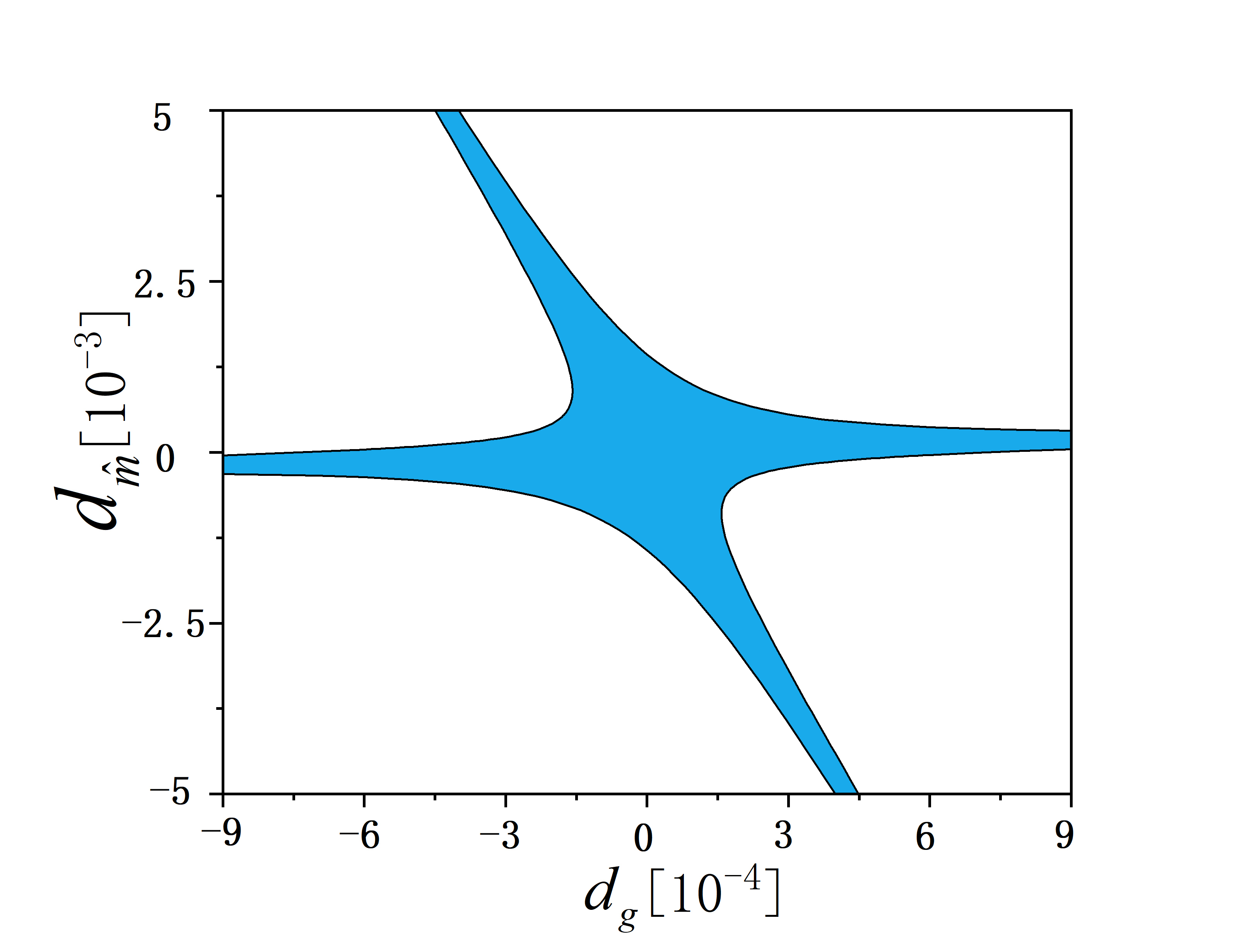}
	\includegraphics[width=0.49\textwidth]{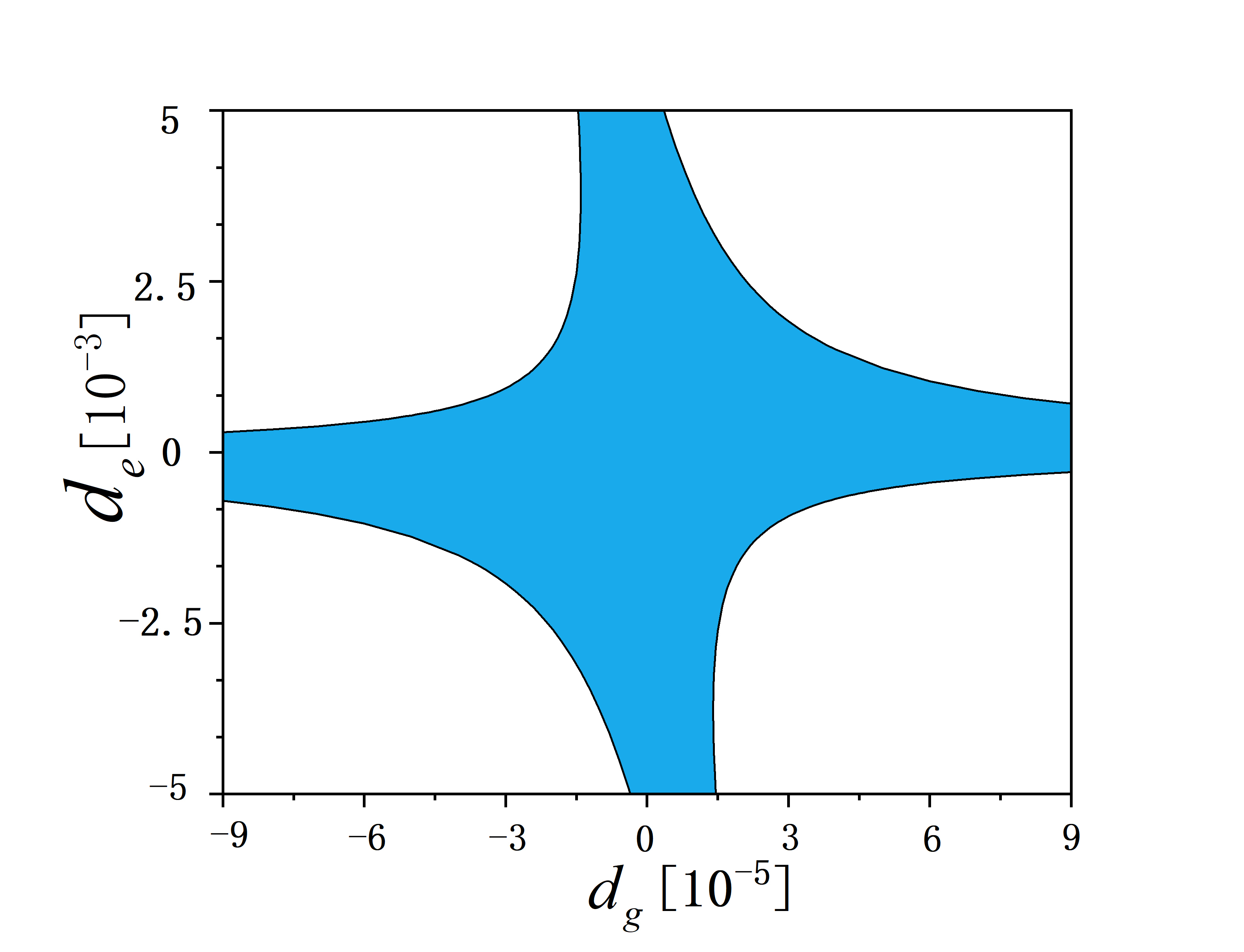}
	\vspace{0.5cm}
	\includegraphics[width=0.49\textwidth]{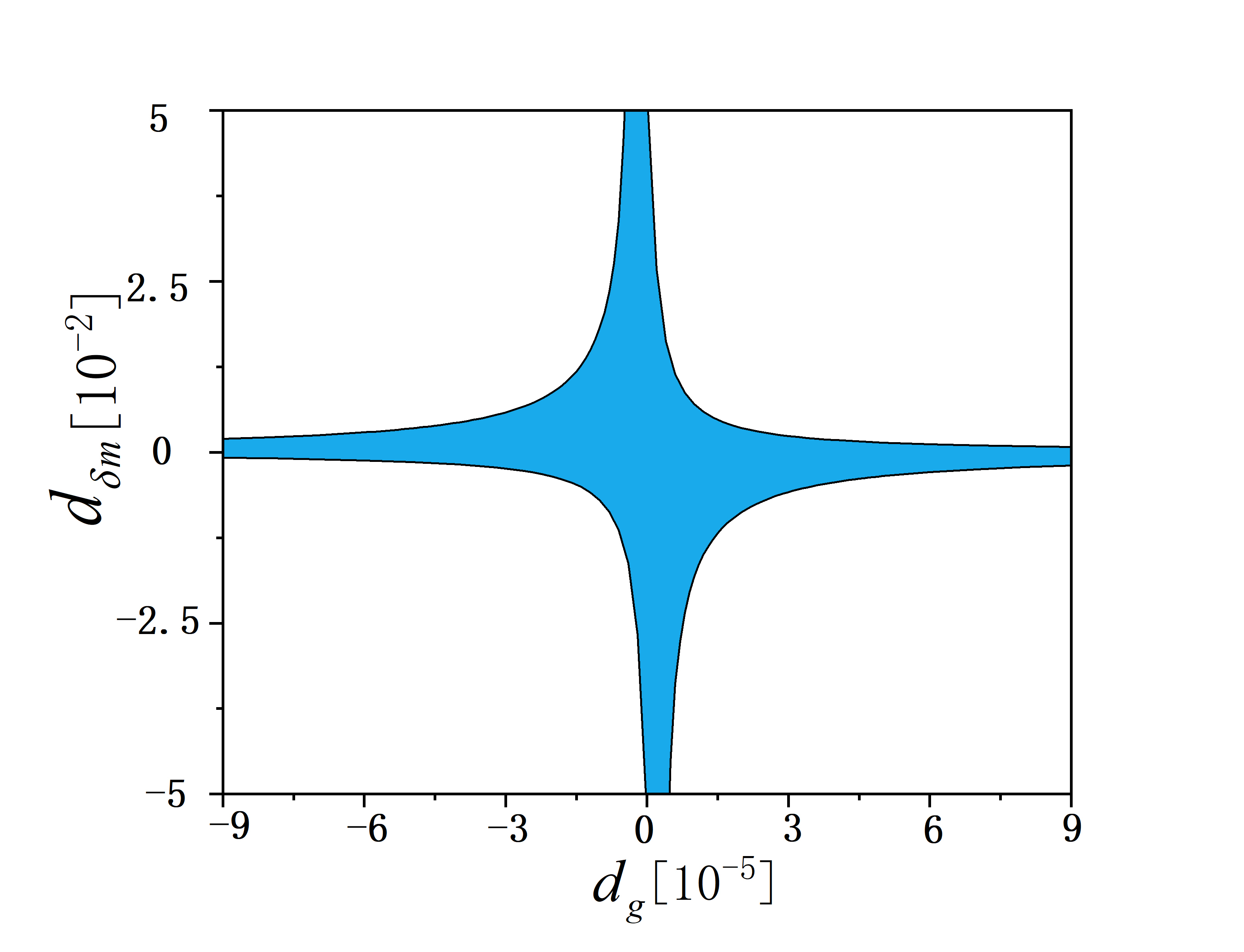}
	\includegraphics[width=0.49\textwidth]{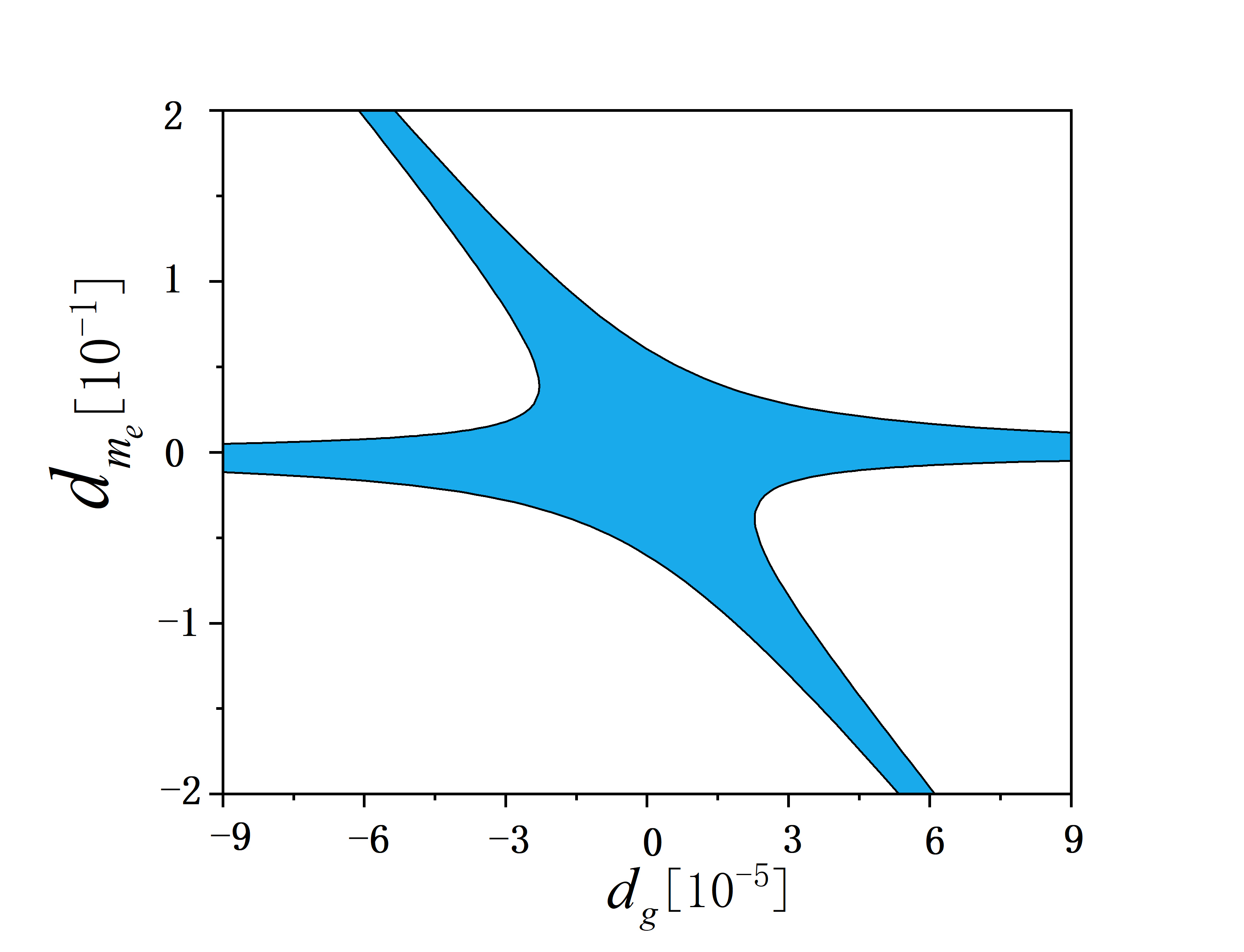} 
	\caption{Constraints on the four pairs ($d_{g}$-$d_{\hat{m}}$, $d_{g}$-$d_{m_{e}}$, $d_{g}$-$d_{e}$ and $d_{g}$-$d_{\delta m}$) set by Asenbaum’s result \cite{PhysRevLett.125.191101}, where the shaded area is the allowed region. We have taken a generic value $10^{-18}$eV for $m_{\varphi}$.}
	\label{fig3}	
\end{figure}

According to Eqs. (\ref{11}) and (\ref{alphaE}), $d_g$ dominates the contribution to $\alpha_A$ and $\alpha_E$, if one assumes that $d_{e}$, $d_{g}$, $d_{m_{e}}$, $d_{\hat{m}}$ and $d_{\delta m}$ are of the same order. To investigate the correlation between $d_g$ and the other four parameters, we assume $d_g$ always nonzero and set one of the other four parameters nonzero each time. Then, we can draw the constraints on the four pairs ($d_{g}$-$d_{\hat{m}}$, $d_{g}$-$d_{m_{e}}$, $d_{g}$-$d_{e}$ and $d_{g}$-$d_{\delta m}$) in Fig. \ref{fig3}, where the result from Ref. \cite{PhysRevLett.125.191101} is used. In Fig. \ref{fig3}, due to the loose constraints we get, we could not see whether there exist correlations between $d_g$ and the other four parameters, or whether $d_g$ is of the same order as the other four parameters.

\section{Conclusion and discussion} \label{conlusion and discussion} 
In this paper, we first generalize the linear coupling scalar DM model to the appearance of a central massive body, such as the Earth. We find that the DM field obtains a local exponential fluctuation term besides the cosmic harmonic oscillation term. Our method can be applied to more general scalar DM models. The case of the quadratic coupling scalar DM model has been studied in Ref. \cite{PhysRevD.98.064051}.

According to Eq. (\ref{effmass}), $m_{{\rm eff}}$ is proportional to the density of the central massive body. For more dense bodies than the Earth, the difference between $m_{{\rm eff}}$ and $m_{\varphi}$ becomes more remarkable. On the other hand, according to Eqs. (\ref{deltaphi1}) and (\ref{deltaphi2}), the $\delta\varphi$ is proportional to the total mass of the central body. Thus, for more massive bodies than the Earth, the local exponential fluctuation term becomes more important.

We then use our solution not only including the cosmic harmonic oscillation term but also the local exponential fluctuation term for the DM field to calculate the DM-induced phase shift in atom interferometers. The resulting phase shift is a sum of a static term and an oscillatory term. Accordingly, for the WEP test with AIs, the E$\ddot{\text{o}}$tv$\ddot{\text{o}}$s parameter $\eta$ is also a sum of a static component and a time-varying component. The two components can be either comparable or one larger than the other depending on the values of $d_i$'s. For current WEP test experiments, the oscillatory component $\eta_{bg}$ makes neglectable contributions in constraining the five DM parameters. For future improved precision of WEP tests, the oscillatory component will become more and more important.

\section*{Acknowledgements}
This work was supported by the National Key Research and Development Program of China under Grant No. 2016YFA0302002, and the Strategic Priority Research Program of the Chinese Academy of Sciences under Grant No. XDB21010100. We thank the referees for their helpful comments and suggestions that significantly polish this work.

\appendix
 \section{Calculation of velocity and position of atoms}\label{AppendixA}
 Solving Eq. (\ref{eom}) is as follows. We first integrate on both sides to get the velocity,	
 \begin{align}
 \dot{z}(t')&=\dot{z}(t_{i})\!-\!g_{0}(t'\!-\!t_{i})\!-\!\alpha_E I(\frac{R_{E}}{\lambda_{\text{eff}}})\frac{GM_{E}}{c^2} \int^{t'}_{t_i}
 \Bigg[\alpha_{A}\Bigg(c^2\!+\!\frac{(\dot{z}^{(0)}(t))^2}{2}\!+\!g_0z^{(0)}(t)\Bigg)\notag \\
 &\cdot\Bigg(\frac{1}{(R_E\!+\!z^{(0)}(t))^2}\!+\!\frac{1}{(R_E+z^{(0)}(t))\lambda_{\text{eff}}}\Bigg)
 +g_0\alpha_E\Bigg(\frac{z^{(0)}(t)}{(R_E+z^{(0)}(t))^2}\notag \\
 &+\frac{1}{R_E+z^{(0)}(t)}(1+\frac{z^{(0)}(t)}{\lambda_{\text{eff}}})\Bigg)e^{-\frac{R_E+z^{(0)}(t)}{\lambda_{\text{eff}}}}\Bigg]dt
 \notag \\
 &+\varphi_{0}\int^{t'}_{t_i}\Bigg[k\Bigg((c^2-\frac{(\dot{z}^{(0)}(t))^2}{2})\alpha_{A}+g_0z^{(0)}(t)(\alpha_{A}+\alpha_E)\Bigg)\sin\Bigg(k(R_{E}\notag \\
 &+z^{(0)}(t))-\omega t+\delta\Bigg)
 -g_0(\alpha_{A}+\alpha_E)\cos\Bigg(k(R_{E}+z^{(0)}(t))-\omega t+\delta\Bigg)\Bigg]dt
 \notag \\
 &-\alpha_{A}\varphi_{0}\Bigg(\cos\Big(k(R_E+z^{(0)}(t'))-\omega t'+\delta\Big)\dot{z}^{(0)}(t')
 -\cos(k(R_E+z^{(0)}_i)\notag \\
 &-\omega t_i+\delta)\dot{z}^{(0)}(t_i)\Bigg)\, .
 \notag \\	
 \label{zdotequ}
 \end{align}
 Here, $z^{(0)}(t)$ denotes the unperturbed atomic trajectory, which is nothing but the freefall trajectory
 \begin{equation}
 z^{(0)}(t)=z_i^{(0)}+v_i^{(0)}(t-t_i)-\frac{1}{2}g_0(t-t_i)^2 \, 
 \end{equation}
 where $t_i$ is the initial time, $z_i^{(0)}$ and $v_i^{(0)}$ are respectively the  initial position and velocity for each segment of the freefall trajectory.
 
 To finish the integration in Eq. (\ref{zdotequ}), we do the following approximation, 
 \begin{align}
 e^{-\frac{R_E+z^{(0)}(t)}{\lambda_{\text{eff}}}} & \simeq  (1-\frac{z^{(0)}(t)}{R_E})\, e^{-\frac{R_E}{\lambda_{\text{eff}}}}
 \notag \\
 \frac{1}{R_E+z^{(0)}(t)}	& \simeq  \frac{1}{R_E}\, (1-\frac{z^{(0)}(t)}{R_E})
 \notag \\
 \frac{1}{(R_E+z^{(0)}(t))^2} & \simeq \frac{1}{R_E^2} \, (1-\frac{2z^{(0)}(t)}{R_E})
 \notag \\
 \end{align}
 Then we can get the result
 \begin{align}
 	\dot{z}(t)=\dot{z}(t_i)-g_0(t-t_i)+\dot{z}_{\rm exp}+\dot{z}_{\rm bg}\, ,	\label{zresult1}
	\end{align}
where
\begin{align}
&\dot{z}_{\rm exp}=-\alpha_Eg_0I(\frac{R_E}{\lambda_{\rm eff}})e^{-\frac{R_E}{\lambda_{\rm eff}}}\frac{1}{R_E\lambda^2_{\rm eff}}\Bigg\{\alpha_A\,(t\!-\!t_i)\Bigg[R_{E}\,{\lambda_{{\rm eff}}}^{2}+{R_{E}}^{2}\lambda_{{\rm eff}}
\notag \\
&\!-\!z_{i}^{(0)}\, \Big(\! 2\,{\lambda_{{\rm eff}}}^{2}\!+\! \big( 2\,R_{E}\!-\!2\,z_{i}^{(0)}
\big) \lambda_{{\rm eff}}\!+\!R_{E}\, \big(\! R_{E}\!-\!z_{i}^{(0)} \big) 
\!\Big) 
\!+\!\frac{1}{10}\big( \lambda_{{\rm eff}}
\!+\!\frac{R_{E}}{2} \big) {g_{0}}^{2}{t_{i}}^{4}
\notag \\
&\!-\!\frac{2}{5}\big(\! tg_{0}\!-\!\frac{5}{4}v_{i}^{(0)} \big)  \big( \!\lambda_{{\rm eff}}\!+\!
\frac{R_{E}}{2} \!\big) g_{0}{t_{i}}^{3}\!+\!\Bigg( \!\frac{3}{5} \big(\! \lambda_{{\rm eff}}\!+\!\frac{R_{E}}{2} \!\big) {g_{0}}^{2}{t
}^{2}
\!-\!\frac{3}{2}v_{i}^{(0)} \big( \lambda_{{\rm eff}}\!+\!\frac{R_{E}}{2} \big) g_{0}t
\notag \\
&\!+\!\frac{1}{3}g_{0}{\lambda_{{\rm eff}}}^{2}\!+\! \Big(\!\big(\! \frac{R_{E}}{3}\!-\!\frac{2}{3}z_
{i}^{(0)} \!\big) g_{0}
\!+\!\frac{2}{3}\,{v_{i}^{(0)}}^{2} \!\Big) \lambda_{{\rm eff}}
\!+\!\frac{1}{6}
\Big(\!\big( R_{E}\!-\!2z_{i}^{(0)} \big) g_{0}\!+\!2{v_{i}^{(0)}}^{2} \!\Big) R_
{E} \!\Bigg) {t_{i}}^{2}\notag \\
&\!+\!\Bigg(\!-\!\frac{2}{5} \big( \lambda_{{\rm eff}}\!+\!\frac{R_{E}}{2} \big) {g_{0}}^{2}{
	t}^{3}
\!+\!\frac{3}{2}v_{i}^{(0)} \big( \lambda_{{\rm eff}}\!+\!\frac{R_{E}}{2} \big) g_{0}
{t}^{2}\!+\! \bigg( \!-\frac{2}{3}\,g_{0}{\lambda_{{\rm eff}}}^{2}+ \Big( 
\big( \frac{4}{3}z_{i}^{(0)}
\notag \\
&-\frac{2}{3}R_{E} \big) g_{0}-\frac{4}{3}{v_{i}^{(0)}}^{2} \Big) 
\lambda_{{\rm eff}}
-\frac{1}{3} \Big(\!\big( R_{E}\!-\!2z_{i}^{(0)} \big) g_{0}\!+\!
2{v_{i}^{(0)}}^{2} \Big) R_{E} \!\bigg) t\!+\!v_{i}^{(0)}\Big( {\lambda_{{\rm 
			eff}}}^{2}
\notag\\
&
+ \big( R_{E}-2z_{i}^{(0)} \big) \lambda_{{\rm eff}}
+\frac{1}{2}R_{
	E} \big( R_{E}
-2\,z_{i}^{(0)} \big)  \Big)  \Bigg) t_{i}+\frac{1}{10}\, \big( \lambda_{{\rm eff}}+\frac{R_{E}}{2} \big) {g_{0}}^{2}{t}^{4}
\notag\\
&\!-\!\frac{1}{2}\,v_{i}^{(0)}\, \big( \lambda_{{\rm eff}}\!+\!\frac{R_{E}}{2} \big) g_{0}\,{t}^{3}\!+\!
\bigg( \frac{1}{3}\,g_{0}\,{\lambda_{{\rm eff}}}^{2}
\!+\! \Big(  \big( \frac{R_{E}}{3}-
\frac{2}{3}\,z_{i}^{(0)} \big) g_{0}\!+\!\frac{2}{3}\,{v_{i}^{(0)}}^{2} \Big) \lambda_{{\rm eff}}
\notag\\
&\!+\!\frac{1}{6}\, \Big(  \big( R_{E}-2\,z_{i}^{(0)} \big) g_{0}\!+\!2\,{v_{i}^{(0)}}^{2}
\Big) R_{E} \bigg) {t}^{2}
-v_{i}^{(0)}\, \bigg( {\lambda_{{\rm eff}}}^{2
}+ \big( R_{E}-2\,z_{i}^{(0)} \big) \lambda_{{\rm eff}}
\notag\\
&
+\frac{1}{2}\,R_{E}\,
\big( R_{E}-2\,z_{i}^{(0)} \big)  \bigg) t
\Bigg]
\notag \\
&+\frac{\alpha_A\big(t-t_i\big)}{12c^2}\,\big(2\,g_{0}\,z_{i}^{(0)}+{v_{i}^{(0)}}^{2}\big)\Bigg[
6\,{R_{E}}^{2}\lambda_{{\rm eff}}+6\,R_{E}\,{\lambda_{{\rm eff}}}^{2}-6\,z_{i}^{(0)}\, \Big( 2\,{\lambda_{{\rm eff}}}^{2}\notag \\
&+ \big( 2\,R_{E}-2\,z_
{i} ^{(0)}\big) \lambda_{{\rm eff}}+R_{E}\, \big( R_{E}-z_{i}^{(0)} \big) 
\Big) 
+\frac{3}{10}\, \left( t-t_{i} \right) ^{4} \big( 2\,\lambda_{{\rm eff}}
+R_{E}
\big) {g_{0}}^{2}
\notag\\
&\!+\! \left( t\!-\!t_{i} \right) ^{2} \bigg( \frac{3}{2}\,v_{i}^{(0)}\,
\left( 2\,\lambda_{{\rm eff}}\!+\!R_{E} \right) t_{i}\!-\!\frac{3}{2}\,v_{i}^{(0)}\,
\left( 2\,\lambda_{{\rm eff}}\!+\!R_{E} \right) t\!-\! \big( 2\,R_{E}
\!+\!4\,
\lambda_{{\rm eff}} \big) z_{i}^{(0)}
\notag \\
&\!+\!{R_{E}}^{2}\!+\!2\,R_{E}\,\lambda_{{\rm 
		eff}}\!+\!2\,{\lambda_{{\rm eff}}}^{2} \!\bigg) g_{0}\!+\!2\,{v_{i}^{(0)}}^{2} \left( 2\,\lambda_{{\rm eff}}\!+\!R_{E} \!\right) {t_{i}}^{2}
\!+\!3\,v_{i}^{(0)}\, \bigg(\!\! \!-\!\frac{4}{3}\,v_{i}^{(0)}\, \big(\! 2\,\lambda_{{\rm eff}}
\notag\\
&\!+\!R_{E}
\big) t\!+\! \left(\! -\!2\,R_{E}-4\,\lambda_{{\rm eff}} \right) z_{i}^{(0)}\!+\!{R_{
		E}}^{2}\!+\!2\,R_{E}\,\lambda_{{\rm eff}}+2\,{\lambda_{{\rm eff}}}^{2}
\bigg) t_{i}+2\,{v_{i}^{(0)}}^{2} \big( 2\,\lambda_{{\rm eff}}
\notag\\
&+R_{E}
\big) {t}^{2}-3\,v_{i}^{(0)}\, \bigg(  \left( -2\,R_{E}-4\,\lambda_{{\rm 
		eff}} \right) z_{i}^{(0)}+{R_{E}}^{2}+2\,R_{E}\,\lambda_{{\rm eff}}+2\,{
	\lambda_{{\rm eff}}}^{2} \bigg) t
\Bigg]
\notag\\
&\!-\!\frac{\alpha_Eg_0\,\big(t-t_i\big)}{20c^2}\Bigg[20\,{R_{E}}^{2}{\lambda_{{
			\rm eff}}}^{2}\!-\!40R_{E}z_{i}^{(0)}{\lambda_{{\rm eff}}}^{2}\!-\!40\,{R_{E}}^{2}z_{i}^{(0)}\lambda_{{\rm eff}}+20{R_{E}}^{2}{z_{i}^{(0)}}^{2}
\notag \\
&\!-\!20R_{E}{z_{i}^{(0)}}^{3}\!+\!60R_{E}{z_{i}^{(0)}}^{
	2}\lambda_{{\rm eff}}\!-\!40{z_{i}^{(0)}}^{3}\lambda_{{\rm eff}}\!+\!40{z_{i}^{(0)}}^{
	2}{\lambda_{{\rm eff}}}^{2}
\!+\!\frac{5}{14}\left( t-t_{i} \right) ^{6} \big( 2\lambda_{{\rm eff}}\notag \\
&\!+\!R_{E}
\big) {g_{0}}^{3}
\!+\!\left( t\!-\!t_{i} \right) ^{4} \big( \!-\!\frac{5}{2}v_{i}^{(0)}t\!+\!\frac{5}{2}v_{i}^{(0)}\,t_{i}+R
_{E}\!-\!3\,z_{i}^{(0)}\!+\!\lambda_{{\rm eff}} \big)  \left( 2\,\lambda_{{\rm eff
	}}\!+\!R_{E} \right) {g_{0}}^{2}
	\notag \\
&	\!-\!5\,g_{0}\left( t\!-\!t_{i}
	\right) ^{2} \bigg( \!-\!\frac{6}{5}{v_{i}^{(0)}}^{2} \left( 2\,\lambda_{{\rm eff}}+R_{E}
	\right) {t_{i}}^{2}\!-\!v_{i}^{(0)}\big( 2\,\lambda_{{\rm eff}}\!+\!R_{E}
	\big)  \big(\! -\!{\frac {12\,v_{i}^{(0)}t}{5}}\!+\!R_{E}
	\notag \\
	&\!-\!3z_{i}^{(0)}\!+\!\lambda_{{
			\rm eff}} \!\big) t_{i}\!-\!\frac{6}{5}\,{v_{i}^{(0)}}^{2} \left( 2\lambda_{{\rm eff}}
	+R_{E} \!\right) {t}^{2}\!+\!v_{i}^{(0)}\left( 2\lambda_{{\rm eff}}\!+\!R_{E}
\!	\right)  \!\left(\! R_{E}\!-\!3z_{i}^{(0)}\!+\!\lambda_{{\rm eff}} \!\right) t
	\notag \\
	&\!+\! \big( 
	\!-\!\frac{4}{3}\,R_{E}\!+\!\frac{8}{3}\,z_{i}^{(0)} \big) {\lambda_{{\rm eff}}}^{2}+ \big( -\frac{4}{3}
	\,{R_{E}}^{2}+4\,R_{E}\,z_{i}^{(0)}-4\,{z_{i}^{(0)}}^{2} \big) \lambda_{{\rm eff
		}}+\frac{4}{3}\,{R_{E}}^{2}z_{i}^{(0)}
	\notag \\	
	&-2\,R_{E}\,{z_{i}^{(0)}}^{2} \bigg)+5\,{v_{i}^{(0)}}^{3} \left( 2\,\lambda_{{\rm eff}}+R_{E} \right) {t_{i}}^{3}+\frac{20}{3}{v_{i}^{(0)}}^{2} \big( -\frac{9}{4}\,v_{i}^{(0)}\,t+R_{E}-3\,z_{i}^{(0)}\notag \\
	&+\lambda_{{\rm eff}}
	\big)  
\big( 2\,\lambda_{{\rm eff}}+R_{E} \big) {t_{i}}^{2}-\frac{40}{3}v_{i}^{(0)}\, \bigg( -\frac {9}{8}{v_{i}^{(0)}}^{2} \big( 2\,\lambda_{{\rm eff}}+R
	_{E} \big) {t}^{2}+v_{i}^{(0)}\, \big( 2\,\lambda_{{\rm eff}}\notag \\
	&+R_{E}
	\big)  \big( R_{E}\!-\!3\,z_{i}^{(0)}\!+\!\lambda_{{\rm eff}} \big) t+ \big( 
	\!-\!\frac{3}{2}\,R_{E}\!+\!3\,z_{i}^{(0)} \big) {\lambda_{{\rm eff}}}^{2}\!+\! \big( -\frac{3}{2}\,{
		R_{E}}^{2}\!+\!\frac{9}{2}\,R_{E}\,z_{i}^{(0)}
\notag \\
&-\frac{9}{2}\,{z_{i}^{(0)}}^{2} \big) \lambda_{{\rm 
			eff}}
	+\frac{3}{2}\,R_{E}\,z_{i}^{(0)}\, \big( R_{E}-\frac{3}{2}\,z_{i}^{(0)} \big)  \bigg) t_{
		i}-5\,{v_{i}^{(0)}}^{3} \left( 2\,\lambda_{{\rm eff}}+R_{E} \right) {t}^{3}
		\notag \\
	&+\frac{20}{3}{v_{i}^{(0)}}^{2} \big( 2\,\lambda_{{\rm eff}}\!+\!R_{E} \big)  \big( R_{E}
-3\,z_{i}^{(0)}\!+\!\lambda_{{\rm eff}} \big) {t}^{2}\!+\!20\,v_{i}^{(0)}\, \bigg(  \left( \!-\!R_{E}\!+\!2\,z_{i}^{(0)} \right) {\lambda_{{\rm eff}
		}}^{2}	\notag \\
		&	+ \left( -{R_{E}}^{2}+3\,R_{E}\,z_{i}^{(0)}-3\,{z_{i}^{(0)}}^{2} \right) 
		\lambda_{{\rm eff}}
		+R_{E}\,z_{i}^{(0)}\, \big( R_{E}-\frac{3}{2}\,z_{i}^{(0)} \big) 
		\bigg) t
\Bigg]\Bigg\}
\end{align}
and
\begin{align}
	&\dot{z}_{\rm bg}=\frac {k\varphi_{0}\,\alpha_{A}\,{c}^{2}}{{\omega}^{3}}\Bigg[\frac{1}{2}\, \bigg(  \Big( g_{0}\, \left( t-t_{i} \right) ^{2}+ \left( -2\,t
	+2\,t_{i} \right) v_{i}^{(0)}-2\,z_{i}^{(0)} \Big) {\omega}^{2}-2\,g_{0}
	\bigg) k
	\notag \\
&\cdot	\sin \left( R_{E}\,k-\omega\,t+\delta \right) +\bigg( \omega-k \Big( g_{0}\, \left( t-t_{i} \right) -v_{i}^{(0)} \Big) 
\bigg) \omega\,\cos \left( R_{E}\,k-\omega\,t+\delta \right)
\notag \\
&\!-\! \left( k\omega v_{i}^{(0)}\!+\!{\omega}^{2} \right) \cos \left( R_{E}k\!-\!
\omega t_{i}+\delta \right) \!+\!k \left( {\omega}^{2
}z_{i}^{(0)}+g_{0} \right) \sin \left( R_{E}k-\omega\,t_{i}+\delta \right) 
\Bigg]
\notag \\
&-\frac{3\alpha_A\varphi_0}{2\omega^5}\Bigg[\Bigg(\bigg( \frac{1}{3}\, \Big(  \left( t-t_{i} \right) ^{2}{g_{0}}^{2}+ \big( 
\left( -2\,t+2\,t_{i} \right) v_{i}^{(0)}-z_{i}^{(0)} \big) g_{0}+\frac{1}{2}\,{v_{i}^{(0)}}^
{2} \Big)  
\notag \\
&\cdot\left( g_{0}\left( t-t_{i} \right) ^{2}+ \left( -2t+
2t_{i} \right) v_{i}^{(0)}-2z_{i}^{(0)} \right) {k}^{2}-\frac{2}{3}g_{0} \bigg) {
	\omega}^{4}+2g_{0} \Big( g_{0} \big( t-t_{i} \big)
\notag \\
&
 \!-\!v_{i}^{(0)} \Big) k{
	\omega}^{3}\!-\!4g_{0}\,{k}^{2} \bigg(  \left( t\!-\!t_{i} \right) ^{2}{g_{0
	}}^{2}\!+\! \Big(  \left( \!-\!2t\!+\!2\,t_{i} \right) v_{i}^{(0)}\!-\!\frac{z_{i}^{(0)}}{2} \Big) g_
	{0}+\frac{3}{4}{v_{i}^{(0)}}^{2} \bigg) {\omega}^{2}
	\notag \\
	&\!+\!8{g_{0}}^{3}{k}^{2}
\Bigg)\sin \left( R_Ek\!-\!\omega\,t\!+\!\delta \right)\!+\!\Bigg(\bigg(  \left( t\!-\!t_{i} \right) ^{2}{g_{0}}^{2}\!+\! \Big(  \left( \!-\!2t\!+\!
2t_{i} \right) v_{i}^{(0)}\!-\!\frac{4}{3}z_{i}^{(0)} \Big) g_{0}
\notag \\
&\!+\!\frac{1}{3}{v_{i}^{(0)}}^{2}
\bigg) {\omega}^{3}\!-\!\frac{4}{3}\Big( \!g_{0} \left( t\!-\!t_{i} \right) \!-\!v_{i}^{(0)} \Big)\bigg(\! 
\left(\! t\!-\!t_{i} \!\right) ^{2}{g_{0}}^{2}\!+\! \Big(\!  \left( -2t\!+\!2t_{i}
\right) v_{i}^{(0)}\!-\!\frac{3}{2}z_{i}^{(0)} \Big) g_{0}
\notag \\
&\!+\!\frac{1}{4}{v_{i}^{(0)}}^{2} \bigg) k{
	\omega}^{2}\!-\!2{g_{0}}^{2}\omega+8{g_{0}}^{2} \Big( g_{0} \left( 
t\!-\!t_{i} \right)\! -\!v_{i}^{(0)} \Big) k
\Bigg)k\omega \cos \left( R_{E}k-\omega\,t+\delta \right)
\notag \\
&+\Bigg(  \bigg( \frac{1}{3}z_{i}^{(0)} \left( -2g_{0}z_{i}^{(0)}+{v_{i}^{(0)}}^{2}
\right) {k}^{2}+\frac{2}{3}g_{0} \bigg) {\omega}^{4}+2v_{i}^{(0)}g_{0}\,k{
	\omega}^{3}+ \big( -2{g_{0}}^{2}z_{i}^{(0)}
\notag \\
&\!+\!3g_{0}{v_{i}^{(0)}}^{2}
\big) {k}^{2}{\omega}^{2}\!-\!8{g_{0}}^{3}{k}^{2} \Bigg) \sin
\left( R_{E}k\!-\!\omega\,t_{i}\!+\!\delta \right)\!-\!\frac{1}{3} \Bigg(  \left( \!-\!4g_{0}z_{i}^{(0)}+{v_{i}^{(0)}}^{2} \right) {\omega}^{3
}
\notag \\
&\!+\!kv_{i}^{(0)}\left( \!-\!6g_{0}z_{i}^{(0)}\!+\!{v_{i}^{(0)}}^{2} \right) {\omega}^{2}\!-\!6
{g_{0}}^{2}\omega\!-\!24v_{i}^{(0)}{g_{0}}^{2}k \Bigg) k\omega \cos
\left( R_{E}k\!-\!\omega\,t_{i}\!+\!\delta \right) 
\Bigg]
\notag \\
&+\frac{\alpha_E\varphi_0g_0}{\omega^5}\Bigg[\Bigg( -\frac{1}{4}\bigg( 2+ \Big( g_{0}\left( t-t_{i} \right) ^{2}-2
tv_{i}^{(0)}+2v_{i}^{(0)}t_{i}-2z_{i}^{(0)} \Big) k \bigg)  \bigg( -2
\notag \\
&+\left( g_{0} \left( t\!-\!t_{i} \right) ^{2}\!-\!2tv_{i}^{(0)}+2v_{i}^{(0)}t_{i}\!-\!
2z_{i}^{(0)}\right) k \bigg) {\omega}^{4}-2 \Big( g_{0} \left( t-t_
{i} \right) -v_{i}^{(0)} \Big) k{\omega}^{3}
\notag \\
&\!+\!3\Big(  \left( t\!-\!t_{i}
\right) ^{2}{g_{0}}^{2}\!+\! \big( \!-\!2tv_{i}^{(0)}\!+\!2v_{i}^{(0)}t_{i}\!-\!\frac{2}{3}z_{i}^{(0)}
\big) g_{0}\!+\!\frac{2}{3}{v_{i}^{(0)}}^{2} \Big) {k}^{2}{\omega}^{2}\!-\!6{g_{0}}
^{2}{k}^{2} \Bigg) 
\notag \\
&\cdot\sin \left( R_{E}\,k-\omega\,t+\delta \right) + \Bigg(  \Big( -g_{0}\, \left( t-t_{i} \right) ^{2}+2\,tv_{i}^{(0)}
-2\,v_{i}^{(0)}\,t_{i}+2\,z_{i}^{(0)} \Big) {\omega}^{3}
\notag \\
&+ \Big( g_{0}\, \left( 
t-t_{i} \right) -v_{i}^{(0)} \Big)  \left( g_{0}\, \left( t-t_{i} \right) 
^{2}-2\,tv_{i}^{(0)}+2\,v_{i}^{(0)}\,t_{i}-2\,z_{i}^{(0)} \right) k{\omega}^{2}+2\,g_{0}
\,\omega
\notag \\
&\!-\!6g_{0} \Big( g_{0} \left( t\!-\!t_{i} \right)\! -\!v_{i}^{(0)}
\Big) k \!\Bigg) \!k\omega\cos \left( R_{E}k\!-\!\omega\,t+\delta \right)\!+\!\Bigg(\!\!\!  \left( {k}^{2}{z_{i}^{(0)}}^{2}\!-\!1 \right) {\omega}^{4}\!-\!2v_{i}^{(0)}k{
	\omega}^{3}
\notag \\
&-2{k}^{2} \left( -g_{0}z_{i}^{(0)}+{v_{i}^{(0)}}^{2} \right) {
	\omega}^{2}+6{g_{0}}^{2}{k}^{2} \Bigg) \sin \left( R_{E}\,k-\omega
t_{i}+\delta \right) - \Bigg( 2v_{i}^{(0)}k{\omega}^{2}z_{i}^{(0)}
\notag \\
&
+2\,{\omega}^
{3}z_{i}^{(0)}
+6v_{i}^{(0)}\,g_{0}\,k+2\,g_{0}\,\omega \Bigg) k\omega\cos \left( R_{E}\,k-\omega\,t_{i}+\delta
\right)
\Bigg]
\notag \\
&-\alpha_{A}\,\varphi_{0}\, \Bigg( \cos \left( R_{E}\,k-\omega\,t+
\delta \right) -\sin \left( R_{E}\,k-\omega\,t+\delta \right) k
\Big( z_{i}^{(0)}+v_{i}^{(0)}\, \left( t-t_{i} \right) 
\notag \\
&-\frac{1}{2}\,g_{0}\, \left( t-t_
{i} \right) ^{2} \Big)  \Bigg)  \Big( v_{i}^{(0)}-g_{0}\, \left( t-t_{i}
\right)  \Big) +\alpha_{A}\,\varphi_{0}\, \Bigg( \cos \left( R_{E}
\,k-\omega\,t_{i}+\delta \right) 
\notag \\
&-\sin \left( R_{E}\,k-\omega\,t_{i}+
\delta \right) kz_{i}^{(0)} \Bigg) v_{i}^{(0)}\, .
\end{align}
It is clear that $\dot{z}_{\rm exp}$ and $\dot{z}_{\rm bg}$ denote effects from the exponential term and the oscillation background term of the DM field, respectively.

	For later use, we give the following velocities. At the time of applying the $\pi$-pulse, the velocity for atoms in the lower arm is $\dot{z}_{1l}\equiv\dot{z}(t)\vert_{t=T}$ with $t_i=0$ and $\dot{z}(t_{i})=v_L$, while the velocity for atoms in the upper arm is $\dot{z}_{1u}\equiv\dot{z}(t)\vert_{t=T}$ with $t_i=0$ and $\dot{z}(t_{i})=v_L+v_R(\varphi)$. At the time of applying the second $\frac{\pi}{2}$-pulse, $\dot{z}_{2l}\equiv\dot{z}(t)\vert_{t=2T}$ with $t_i=T$ and $\dot{z}(t_{i})=\dot{z}_{1l}+v_R(\varphi)$, while $\dot{z}_{2u}\equiv\dot{z}(t)\vert_{t=2T}$ with $t_i=T$ and $\dot{z}(t_{i})=\dot{z}_{1u}-v_R(\varphi)$.
	
	Next, we do the time integration on Eq. (\ref{zresult1}) to get the solution for the trajectory
	\begin{align}
	z(t)= z(t_{i})+\int^{t}_{t_i}\dot{z}(t') dt' \, .
	\label{zresult2}
	\end{align}
	For later use, we  give the following positions. At the time of applying the $\pi$-pulse, the position for atoms in the lower arm is $z_{1l}\equiv z(t)\vert_{t=T}$ with $t_{i}=0$, $z(t_{i})=0$ and $\dot{z}(t_{i})=v_L$, while the position for atoms in the upper arm is $z_{1u}\equiv z(t)\vert_{t=T}$ with $t_{i}=0$, $z(t_{i})=0$ and $\dot{z}(t_{i})=v_L+v_R(\varphi)$. At the time of applying the second $\frac{\pi}{2}$-pulse, $z_{2l}\equiv z(t)\vert_{t=2T}$ with $t_{i}=T$, $z(t_{i})=z_{1l}$ and $\dot{z}(t_{i})=\dot{z}_{1l}+v_R(\varphi)$, while $z_{2u}\equiv z(t)\vert_{t=2T}$ with $t_{i}=T$, $z(t_{i})=z_{1u}$ and $\dot{z}(t_{i})=\dot{z}_{1u}-v_R(\varphi)$.
\section{Calculation of the DM-induced phase shift in AI experiments}\label{AppendixB}
The total phase shift can be written as a sum of three components \cite{PhysRevD.78.042003}, the propagation phase shift, the laser phase shift, and the separation phase shift, 
\begin{equation}
\Delta \phi=\Delta\phi_{prop}+\Delta\phi_{laser}+\Delta\phi_{sep} \, .
\end{equation}

For each segment of the atomic trajectory, the atom accumulates a propagation phase
\begin{align}
\phi_{prop}=\int^{t_f}_{t_i}L\, dt \, , 
\end{align}
where $t_f$ is the final time for each segment, and $L$ is the Lagrangian (\ref{GAI3}).
The propagation phase shift $\Delta\phi_{prop}$ is the difference in the propagation phase between the two arms,
\begin{align}
\Delta\phi_{prop}=\sum_{{\rm upper}}\phi_{prop}-\sum_{{\rm lower}}\phi_{prop} \, .\label{GAI6} 
\end{align}

The laser phase shift comes from the interaction of laser pulses with atoms. At each interaction point, the laser field transfers its phase to the atom. Then, $\Delta\phi_{laser}$ is the difference in the accumulated laser phase between the upper and lower arms
\begin{align}
\Delta\phi_{laser}&=\sum_{{\rm upper}}\phi_{laser} - \sum_{{\rm lower}}\phi_{laser} 
\notag \\
&=c\int_{0}^{\frac{z_{i}}{c}}k_{eff}(t)dt-c\int_{T}^{T+\frac{z_{1u}}{c}}k_{eff}(t)dt
\notag \\
&-c\int_{T}^{T+\frac{z_{1l}}{c}}k_{eff}(t)dt+c\int_{2T}^{2T+\frac{z_{2l}}{c}}k_{eff}(t)dt\, ,
\end{align}
where $z_i$ is the initial position of atoms at the time of applying the first $\frac{\pi}{2}$-pulse.

Since the two arms do not exactly intersect at the final laser pulse, then the separation phase shift $\Delta\phi_{sep}$ appears. 
\begin{align}
\Delta\phi_{sep}=\frac{m_{A}}{2\hbar}(\dot{z}_{2u}-v_R+\dot{z}_{2l})(z_{2l}-z_{2u}) \, .
\end{align}

With Eqs. (\ref{zresult1}) and (\ref{zresult2}), we can calculate $\phi_{prop}$ along each segment of the lower and upper arms, and thus compute $\Delta\phi_{prop}$. Similarly, $\Delta\phi_{laser}$ and $\Delta\phi_{sep}$ can also be computed. Summing them together, one can get the final result for the DM-induced phase shift. We find that $\Delta \phi$ consists of a static component $\Delta\phi_{\delta\varphi}$, an oscillatory component
$\Delta\phi_{bg}$, and the well-known term $-g_{0}T^2k_{\text{eff}}$, 
\begin{align}
\Delta \phi	=-g_{0}T^2k_{\text{eff}}+\Delta\phi_{\delta\varphi}+\Delta\phi_{bg} \, .
\label{phaseresult1}
\end{align}
The first term is the known phase shift for atoms in freefall, where ${\textbf k}_{{\rm eff}}$ has been  taken to be parallel to ${\textbf g}_0$.
The $\delta\varphi$-contribution to $\Delta \phi$ is given by
\begin{align}
\Delta\phi_{\delta\varphi}&=\!-\!g_{0}T^2k_{\text{eff}}\Bigg[\Bigg(\frac{\frac{7}{6}g_{0}T^2-(2v_{L}+v_{R})T}{\lambda_{\text{eff}}}\!+\!(1+\frac{R_{E}}{\lambda_{\text{eff}}})\!+\!\frac{ v_{L}(v_{R}+v_{L})}{2c^2}\Bigg)\alpha_{A}
\notag \\
&+\Bigg(\frac{g_{0}(2\,v_{L}\!+\!v_{R})T\!-\!g_{0}R_{E}}{c^2}\!-\!\frac{7\,g^2_{0}\,{T}^{2}}{6c^2}\Bigg)\alpha_{E}
\!+\!\frac{1}{\lambda_{\text{eff}}c^2}\Bigg(\Big(\big(\frac{7}{12}v_{L}(v_{L}\!+\!v_{R})\notag \\
&\!-\!\frac{1}{12}v_{R}^2\big)g_{0}T^2\!-\!v_{L}(v_{L}\!+\!v_{R})\big((v_{L}\!+\!\frac{1}{2}v_{R})T
-\frac{1}{2}R_{E}\big)\Big)\alpha_{A}
+\Big(-{\frac {31}{20}}{T}^{3}{g_{0}}^{2}\notag \\
&+\frac {9}{2}\, \big( v_{L}+\frac {v_{R}}{2}
\big) g_{0}\,{T}^{2}-\big(\,  \frac{7}{6}\,g_{0}\,R_{E}+\frac{7}{2}\,{v_{L}}(v_{L}+\,v_{R})+{v_{R}}^{2}
\big) T+R_{E}\big(2v_{L}\notag \\
&+v_{R}\big)
\Big)
 g_{0}T\alpha_{E}\Bigg)\Bigg]I(\frac{R_{E}}{\lambda_{\text{eff}}})\alpha_{E}e^{-\frac{R_{E}}{\lambda_{\text{eff}}}}
\notag \\
&+g_{0}k_{\text{eff}}T^2\Bigg[\frac{1}{c^2}\Bigg(\frac{17}{2}\,{g_{0}}^{2}{T}^{2}+\frac{1}{2}\, \left( -26\,v_{L}\,T-15\,Tv_{R}+2\,R_{E}
\right) g_{0}+ \big(4 v_{L}\notag \\
&+v_{R} \big)  \left( v_{R}+v_{L}
\right)\Bigg)+\frac{1}{\lambda_{\text{eff}}c^2} \Bigg({\frac {161}{4}\,{g_{0}}^{3}{T}^{4}}
-93\, \left( v_{L}+{\frac {37\,}{62}v_{
		R}} \right) {g_{0}}^{2}{T}^{3}\notag \\
&+\frac{1}{2}\,g_{0}\, \left( 17\,g_{0}\,R_{E}+138\,{v_{L}}^{2}+168\,v_{R}\,v_{L}+49
\,{v_{R}}^{2} \right) {T}^{2}-\frac{1}{2}\, \big( 26\,g_{0}\,R_{E}v_{L}\notag \\
&+15\,g_{0}\,R_{E}v_{R}+32\,{v_{L}}^{3}+60\,
v_{R}\,{v_{L}}^{2}+34\,{v_{R}}^{2}v_{L}+6\,{v_{R}}^{3} \big) T
+\big( 4\,v_{L}\notag \\
&+v_{R} \big) R_{E} \left( v_{R}+v_{L} \right)\Bigg) 
\Bigg]I(\frac{R_{E}}{\lambda_{\text{eff}}})\tilde{d}\alpha_{E}e^{-\frac{R_{E}}{\lambda_{\text{eff}}}}
\notag \\
\label{phaseresult2}
\end{align}
The $\varphi_{bg}$-contribution to $\Delta \phi$ is given by
\begin{align}
\Delta\phi_{bg}&=-k_{\text{eff}}\frac{c^2k\alpha_{A}\varphi_{0}}{\omega^2}\Bigg(\sin(kR_{E}-2\omega T+\delta)-2\sin(kR_{E}-\omega T+\delta)\notag \\
&+\sin(kR_{E}+\delta)\Bigg)	+\alpha_{A}\frac{2g_{0}k_{\text{eff}}T}{\omega}\varphi_{0}\Bigg(\sin(kR_{E}-\omega T+\delta)-\sin\big(kR_{E}\notag \\
&-2\omega T+\delta\big)\Bigg)
+\left( \alpha_{E}+2\alpha_{A}\right) \frac{g_{0}k_{\text{eff}}}{\omega^{2}}\varphi_{0}\Bigg(\cos(kR_{E}+\delta)-2\cos\big(kR_{E}\notag \\
&\!-\!\omega T+\delta\big)\!+\!\cos(kR_{E}-2\omega T+\delta)\Bigg)
\!-\!\alpha_{A}(\frac{k_{\text{eff}}(v_{L}+\frac{v_{R}}{2})}{\omega})\varphi_{0}\Bigg(\sin\big(kR_{E}\notag \\
&+\delta\big)+2\sin(kR_{E}-\omega T+\delta)-\sin(kR_{E}-2\omega T+\delta)\Bigg)
\notag\\
&-k_{\text{eff}}\frac{k\varphi_{0}}{\omega^3}\Bigg[ \big( 4g_{0}T-2v_{L}-v_{R} \big)  \bigg(   
\big( g_{0}T^2-v_{L}T-\frac{v_{R}}{2}T \big) {\omega}^{2}\alpha_{A}-\frac{9}{2}\,g_{0}  
\alpha_{A}\notag \\
&-2\,g_{0}\,\alpha_{E} \bigg)\cos(kR_{E}-2\omega T+\delta) 
-\big( g_{0}T-v_{L}-\frac{1}{2}v_{R} \big)  \bigg(   
\big( g_{0}T^2-2v_{L}T\notag \\
&-v_{R}T \big) {\omega}^{2}\alpha_{A}-18\,g_{0}  
\alpha_{A}-8\,g_{0}\,\alpha_{E} \bigg)\cos(kR_{E}-\omega T+\delta) 	
+g_{0}\big(2\alpha_{E}\notag \\
&+\frac{9}{2}\alpha_{A}\big)
(2v_{L}+v_{R})\cos(kR_{E}+\delta)\Bigg]
+k_{\text{eff}}\frac{k\varphi_{0}}{\omega^4}\Bigg[\Bigg(\bigg( 2g_{0}\,T   \big( 2\,g_{0}T-2\,
v_{L}\notag \\
&-v_{R} \big)\left( \alpha_{E}+3\,\alpha_{A} \right) +  \frac{3}{2}\,v_{L}\, \left( v_{L}+v_{R}
\right)\alpha_{A} 
+\dfrac{1}{2}\,{v_{R}}^{2} \alpha_{A} \bigg) {\omega}^{2}-12\,{g_
	{0}}^{2} \big( \alpha_{A}\notag \\
&\!+\!\frac{\alpha_{E}}{2} \big) 
\Bigg) \sin(kR_{E}-2\omega T+\delta)\!+\!\Bigg(\bigg( 2g_{0}\,T   \left( 2\,
v_{L}\!+\!v_{R}\!-\!g_{0}T \right)\left( \alpha_{E}+3\,\alpha_{A} \right)
\notag \\
& \!-\! 3\,v_{L}\, \left( v_{L}\!+\!v_{R}
\right)\alpha_{A}\! -\!{v_{R}}^{2} \alpha_{A} \bigg) {\omega}^{2}+24\,{g_
	{0}}^{2} \big( \alpha_{A}+\frac{\alpha_{E}}{2} \big) 
\Bigg) \sin\big(kR_{E}-\omega T\notag \\
&\!+\!\delta\big)\!+\!\Bigg(\big(\frac{3}{2}v_{L}(v_{L}\!+\!v_{R})\!+\!\frac{1}{2}v_R^2\big)\omega^2\alpha_{A}
-12g_{0}^2(\alpha_{A}\!+\!\frac{\alpha_{E}}{2})\Bigg)\sin(kR_{E}+\delta)\Bigg]
\notag \\
&-\frac{k_{\text{eff}}T^2k\tilde{d}\varphi_{0}}{ c^2}\Bigg[ \left( 4\,{T}^{2}{g_{0}}^{2}-\left( 8c+4v_{L}+2v_{R} \right) g_{
	0}\,T+4c \left( c+v_{L}+v_{R} \right)  \right) \notag \\
&\cdot\left( Tg_{0}
\!-\!v_{L}\!-\!\frac{1}{2}v_{R}\right) ^{2}\sin \left( kR_{E}\!-\!2\,T\omega+\delta \right)
+\Bigg( -\frac{1}{8}\,{g_{0}}^{3}{T}^{3}+\frac{1}{4}\, \big( c+3\,v_{L}\notag \\
&+\frac{3}{2}\,v_{R}
\big) {g_{0}}^{2}{T}^{2}-g_{0}\,T \bigg( \frac{3}{2}\,{v_{L}}^{2}+ \big( c
+\frac{3}{2}\,v_{R} \big)  \big( v_{L}+\frac{v_{R}}{2} \big)  \bigg) + \bigg( {v
	_{L}}^{2}+v_{L}\,v_{R}
\notag \\
&+\frac{1}{2}\,{v_{R}}^{2} \bigg) c+\frac{1}{2}\,{v_{L}}^{3}+\frac{1}{2}
\, \left( v_{L}+v_{R} \right) ^{3}\Bigg)  \left( g_{0}\,T-2\,c
\right)\sin(kR_{E}-\omega T+\delta) \Bigg]
\notag \\
&-\frac{k_{\text{eff}}T\tilde{d}\varphi_{0}}{ \omega c}\Bigg[ \left( k{g_{0}}^{2}{T}^{2}- \left( v_{L}+\frac{v_{R}}{2} \right) g_{0}\,kT+c\,
\omega\right)  \left( 2\,Tg_{0}-2\,v_{L}-v_{R} \right) \notag \\
&\cdot\cos \left( kR_{E}
-2\omega\,T+\delta \right) 
+\Bigg( -\frac{1}{4}\,k{g_{0}}^{3}{T}^{3}+\left( v_{L}+\frac{v_{R}}{2} \right) {
	g_{0}}^{2}k{T}^{2}- \bigg(  \big( {v_{L}}^{2}\notag \\
&+v_{L}\,v_{R}+\frac{1}{2}\,{v_{R
	}}^{2} \big) k+c\,\omega \bigg) g_{0}\,T- \left( 2 v_{L}+v_{R}
	\right) c\,\omega \Bigg) 
	\cos \left( kR_{E}-T\omega+\delta \right) 
	\Bigg]
	\label{phaseresult3}
	\end{align}	

\bibliography{aidm202011}

\end{document}